\documentclass[mathleft,fleqn,%
]{an}

\usepackage{graphicx}
\usepackage{amsmath}
\usepackage[varg]{txfonts}
\usepackage{epsfig}
\usepackage{times}
\usepackage{rotating}
\usepackage{setspace}
\usepackage{longtable}
\usepackage{supertabular,booktabs}
\usepackage{lscape}
\usepackage[usenames]{xcolor}
\usepackage{natbib}
\bibpunct{(}{)}{;}{a}{}{,}
\newcommand{\kms}{$\rm km\;s^{-1}$}

\newcommand{\msun}{$\rm M_\odot$}
\newcommand{\mas}{mag arcsec$^{-2}$}

\newcommand{\apix}{$\rm \AA\;pixel^{-1}$}

\newcommand{\sigmac}{$\sigma_{\rm c}$}
\newcommand{\sigmare}{$\sigma_{\rm e}$}
\newcommand{\sigmas}{$\sigma_{\star}$}
\newcommand{\sigmag}{$\sigma_{\rm gas}$}
\newcommand{\mbh}{$M_{\bullet}$}
\newcommand{\msigmas}{$M_{\bullet}-\sigma_{\star}$}
\newcommand{\msigmac}{$M_{\bullet}-\sigma_{\rm c}$}

\def\nii{[\ion{N}{ii}]$\lambda\lambda6548,6583$}

\def\niig{[\ion{N}{ii}]$\lambda6583$}
\def\halpha{H$\alpha$}
\def\sii{[\ion{S}{ii}]$\lambda\lambda6716,6731$}

\newcommand{\placetablegalsample}{
\begin{table*}
\begin{small}
\begin{center}
\caption{Properties of the galaxy sample.} 
\begin{footnotesize}
\begin{tabular}{cccccccccc}
\hline
\hline
	\multicolumn{1}{c}{Galaxy} & 
	\multicolumn{1}{c}{Mor. T.} & 
	\multicolumn{1}{c}{Bar} & 
	\multicolumn{1}{c}{Sp. Cl.} &
	\multicolumn{1}{c}{$D$} & 
	\multicolumn{1}{c}{$M^0_B$} &
	\multicolumn{1}{c}{\sigmac} &
	\multicolumn{1}{c}{\sigmare} &
	\multicolumn{1}{c}{\mbh\ [$33^{\circ}$]} &
	\multicolumn{1}{c}{\mbh\ [$81^{\circ}$]} \\
 & & & & [Mpc] & [mag] & [\kms] & [\kms] & [\msun] & [\msun] \\
(1)&(2)&(3)&(4)&(5)&(6)&(7)&(8)&(9)&(10)\\
\hline
NGC~2654 & SBab: sp   & yes & -   & 19.4 & $-19.77$ & $150\pm3$ & $162\pm3$ & 1.9e7 & 4.2e6\\
NGC~3049 & SB(rs)ab   & yes & H   & 23.9 & $-19.12$ & $84\pm9$  & $86\pm10$ & 4.3e6 & 1.2e6\\
NGC~3259 & SAB(rs)bc: & no  & S1  & 24.0 & $-19.40$ & $76\pm7$  & $63\pm6$ & 3.6e6 & 1.0e6\\
NGC~4343 & SA(rs)b:   & no  & -   & 18.1 & $-18.92$ & $110\pm4$ & $154\pm5$ & 6.4e7 & 5.1e7\\
NGC~5141 & S0         & -   & -   & 72.4 & $-20.59$ & $248\pm7$ & $222\pm6$ & 6.3e8 & 2.0e8\\
NGC~5635 & S pec      & no  & -   & 60.1 & $-20.96$ & $238\pm6$ & $216\pm5$ & 1.1e9 & 2.9e8\\
NGC~5713 & SAB(rs)bc pec & yes & H& 28.3 & $-20.68$ & $71\pm5$  & $78\pm6$ & 3.4e7 & 1.4e7\\
\hline
\end{tabular}
\label{tab:sample}
\end{footnotesize}
\end{center}
\emph{Notes}. Col.(1): galaxy name. Col.(2): morphological type from
\citet[][RC3]{deVaucouleurs1991}. Col.(3): presence of the bar
according to the photometric decomposition.  We inferred the presence
of a bar in the edge-on galaxy NGC~2654 from its boxy/peanut bulge.  
NGC~5141 turned out to be an elliptical galaxy.
Col.(4): nuclear spectral class from NED, where H = HII nucleus, L =
LINER, S1 = Seyfert of type 1. Col.(5): distance. The distances were 
obtained as $D = V_{\rm 3K}/H_{0}$, where $V_{\rm 3K}$ is the weighted 
mean recessional velocity corrected to the reference frame of the 
microwave background radiation given in RC3 and adopting $H_{0} = \rm
75~km~s^{-1}~Mpc^{-1}$. Col.(6): absolute total corrected $B$
magnitude obtained from $B^{0}_{\rm T}$ (RC3) and adopted
distance. Col.(7): stellar velocity dispersion within $r_{\rm
  e}/8$ from this paper. Col.(8): stellar effective velocity dispersion 
  within circularized $r_{\rm e}$ from the aperture 
  correction from \citet{Falcon2017}. Col.(9): stringent \mbh\ limit assuming 
  $i= 33^{\circ}$ for the gas disk from this paper. Col.(10): stringent \mbh\ limit 
  assuming $i= 81^{\circ}$ for the gas disk from this paper.
\end{small}
\end{table*}}

\newcommand{\placetablegasptwod}{
\begin{table*}
\begin{small}
\begin{center}
\caption{Structural parameters of the sample galaxies from the
  photometric decomposition of the $i$-band SDSS images.}
\begin{footnotesize}
\begin{tabular}{lccccc}
\hline
	\multicolumn{1}{c}{Parameter} & 
	\multicolumn{1}{c}{NGC~3049} &
	\multicolumn{1}{c}{NGC~3259} &
	\multicolumn{1}{c}{NGC~5141} &
	\multicolumn{1}{c}{NGC~5713} \\
	\multicolumn{1}{c}{(1)} & 
	\multicolumn{1}{c}{(2)} &
	\multicolumn{1}{c}{(3)} &
	\multicolumn{1}{c}{(4)} &
	\multicolumn{1}{c}{(5)}\\
\hline 
\vspace{-3mm} \\
$\mu_{\rm e}$ [\mas] & $19.4\pm0.8$& $19.4\pm0.6$& $21.25\pm0.07$& $19.8\pm0.4$\\
$r_{\rm e}$ [$\rm arcsec$] & $5.2\pm0.1$& $1.06\pm0.02$& $18.0\pm0.2$& $8.7\pm0.2$\\
$n$ & $0.97\pm0.02$ & $1.28\pm0.01$ & $5.58\pm0.03$& $1.93\pm0.02$\\
$q_{\rm bulge}$ & $0.270\pm0.006$& $1.000\pm0.003$& $0.704\pm0.001$& $0.311\pm0.005$\\
${\it PA}_{\rm bulge}$ [$^{\circ}$] & $35.8\pm0.9$& $13.8\pm0.5$& $74.3\pm0.1$& $115.1\pm0.5$\\
$\mu_{\rm 0}$ [\mas] & $21.64\pm0.03$& $18.7\pm0.6$& - & $19.6\pm0.1$\\
$h$ [$\rm arcsec$] & $64.1\pm0.6$& $9.1\pm0.1$& - & $22.0\pm0.1$\\
$h_{\rm out}$ [$\rm arcsec$] & $16.0\pm0.3$ & - & - & - \\
$r_{\rm break}$ [$\rm arcsec$] & $47.4\pm0.7$& - & - & - \\
$q_{\rm disk} $ & $0.596\pm0.002$& $0.599\pm0.002$& - & $0.916\pm0.001$\\
${\it PA}_{\rm disk} $ [$^{\circ}$] & $27.36\pm0.09$& $13.7\pm0.2$& - & $33.15\pm0.05$\\
$\mu_{\rm 0,bar}$ [\mas] & $20.97\pm0.03$& - & - & $19.91\pm0.04$\\
$a_{\rm bar}$ [$\rm arcsec$] & $60.8\pm0.2$ & - & - & $40.05\pm0.07$\\
$q_{\rm bar} $ & $0.182\pm0.001$& - & - & $0.549\pm0.001$\\
${\it PA}_{\rm bar} $ [$^{\circ}$] & $29.74\pm0.06$& - & - & $98.73\pm0.03$\\
${\it L}_{\rm bulge}/L_{\rm T}$ & 0.11 & 0.03 & 1.00 & 0.08 \\
${\it L}_{\rm disk}/L_{\rm T}$ & 0.71 & 0.97 & - & 0.76 \\
${\it L}_{\rm bar}/L_{\rm T}$ & 0.18 & - & - & 0.16 \\
\hline
\end{tabular}
\label{tab:gasp2d}
\end{footnotesize}
\end{center}
\emph{Notes}. ${\it L}_{\rm bulge}/{\it L}_{\rm T}$, ${\it L}_{\rm disk}/{\it L}_{\rm
  T}$, and ${\it L}_{\rm bar}/{\it L}_{\rm T}$ are the bulge-to-total,
disk-to-total, and bar-to-total luminosity ratio,
respectively. 
\end{small}
\end{table*}}

\newcommand{\placetableasiago}{
\begin{table}
\begin{small}
\begin{center}
\caption{Details of the T122/B\&C observations and central stellar
  velocity dispersion of the sample galaxies.}
\begin{footnotesize}
\begin{tabular}{cccccc}
\hline
\hline
	\multicolumn{1}{c}{Galaxy} & 
	\multicolumn{1}{c}{${\it PA}$} & 
	\multicolumn{1}{c}{Exp. T.} &
	\multicolumn{2}{c}{Apert.} &
	\multicolumn{1}{c}{\sigmas} \\
 & [$^{\circ}$] & [h] & [arcsec] & [pc] & [\kms] \\
(1)&(2)&(3)&(4)&(5)&(6) \\
\hline
NGC~2654 & 63 & 3.0 & $3\times 2$ & $282\times 188$ & $144\pm3$ \\
NGC~3049 & 25 & 5.0 & $5\times 2$ & $579\times 232$ & $81\pm9$  \\
NGC~3259 & 20 & 3.5 & $5\times 2$ & $581\times 233$ & $68\pm6$  \\
NGC~4343 &133 & 3.0 & $3\times 2$ & $263\times 175$ & $112\pm4$ \\
NGC~5141 & 80 & 3.0 & $3\times 2$ & $1053\times702$ & $253\pm7$ \\
NGC~5635 & 65 & 3.5 & $3\times 2$ & $874\times 583$ & $235\pm6$ \\
NGC~5713 & 10 & 3.0 & $3\times 2$ & $412\times 274$ & $70\pm5$  \\
\hline
\end{tabular}
\label{tab:asiago}
\end{footnotesize}
\end{center}
\emph{Notes}. Col.(1): galaxy name. Col.(2): position angle of the
slit along the galaxy major axis as given by RC3. Col.(3): total
exposure time. Col.(4): size of the central aperture where we measured
the stellar velocity dispersion. Col.(5): physical size of the central 
aperture where we measured the stellar velocity dispersion. Col.(6): 
central stellar velocity dispersion.
\end{small}
\end{table}}

\newcommand{\placetablehst}{
\begin{table*}
\begin{small}
\begin{center}
\caption{Details of the HST/STIS observations and central gas velocity
  dispersion of the sample galaxies.}
\begin{footnotesize}
\begin{tabular}{cccccccccc}
\hline
\hline
	\multicolumn{1}{c}{Galaxy} & 
	\multicolumn{1}{c}{Prop. Id.}  &
	\multicolumn{1}{c}{${\it PA}$}  &
	\multicolumn{1}{c}{Exp. T.}  &
	\multicolumn{1}{c}{Sp. Range}  &
	\multicolumn{1}{c}{Slit}  &
	\multicolumn{1}{c}{Bin.} & 
	\multicolumn{2}{c}{Apert.} &
	\multicolumn{1}{c}{\sigmag} \\
 & & [$^{\circ}$] & [h] & [$\AA$] & [arcsec] & & [arcsec] & [pc] & [\kms] \\ 
(1)&(2)&(3)&(4)&(5)&(6)&(7)&(8)&(9)&(10)\\
\hline
NGC~2654 & 9046 &  62.0 & 1.25 & $6480\,$--$\,7060$ & 0.1 & $1\times1$ & $0.15\times0.1$ & $14\times 9$ & $71\pm13$\\
NGC~3049 & 7513 &  42.1 & 0.27 & $6300\,$--$\,6870$ & 0.1 & $1\times1$ & $0.15\times0.1$ & $17\times12$ & $31\pm2$\\
NGC~3259 & 8228 &  20.1 & 0.24 & $6480\,$--$\,7060$ & 0.2 & $1\times1$ & $0.25\times0.2$ & $29\times23$ & $23\pm3$\\
NGC~4343 & 9068 & 144.1 & 0.81 & $6300\,$--$\,6870$ & 0.2 & $1\times1$ & $0.25\times0.2$ & $22\times18$ & $79\pm15$\\
NGC~5141 & 8236 &  72.3 & 0.36 & $6480\,$--$\,7060$ & 0.2 & $2\times2$ & $0.30\times0.2$ &$105\times70$ &$142\pm6$\\
NGC~5635 & 7354 &  63.6 & 0.13 & $6480\,$--$\,7060$ & 0.1 & $2\times1$ & $0.15\times0.1$ & $44\times29$ &$333\pm25$\\
NGC~5713 & 8228 &  10.1 & 0.20 & $6480\,$--$\,7060$ & 0.2 & $2\times2$ & $0.30\times0.2$ & $41\times27$ & $49\pm6$\\
\hline
\end{tabular}
\label{tab:hst}
\end{footnotesize}
\end{center}
\emph{Notes}. Col.(1): galaxy name. Col.(2): HST proposal
number. Col.(3): position angle of the slit. Col.(4): total exposure
time. Col.(5): spectral range. Col.(6): size of the slit. Col.(7):
pixel binning. Col.(8): size of the central aperture where the gas
velocity dispersion was measured. Col.(9): physical size of the
central aperture where the gas velocity dispersion was
measured. Col.(10): central gas velocity dispersion.
\end{small}
\end{table*}}

\newcommand{\placefiguredecone}{
\begin{figure*}
\begin{small}
\begin{center}
\includegraphics[scale= 0.56,angle=-90]{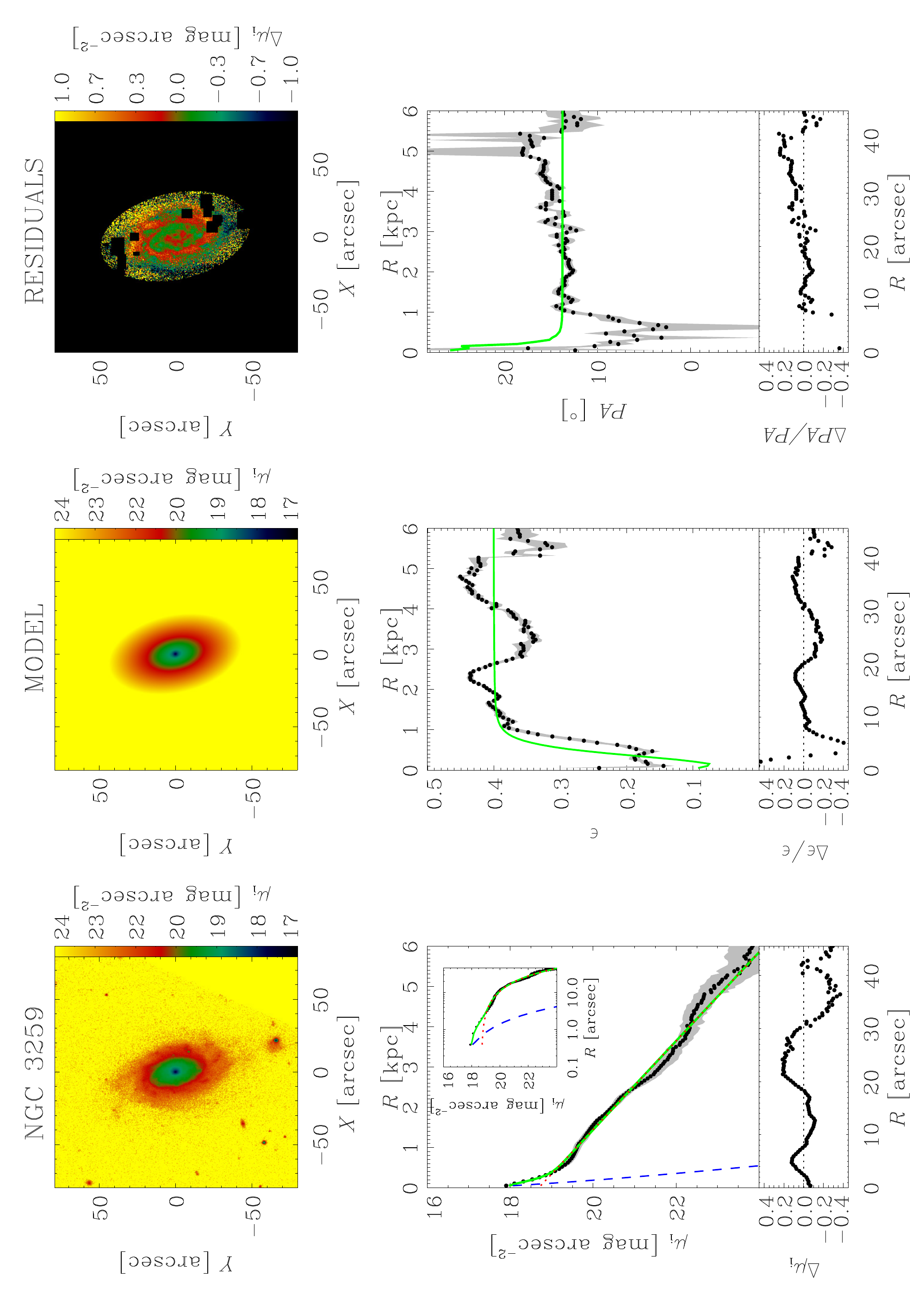}
\caption{Two-dimensional photometric decomposition of the $i$-band
  SDSS image of NGC~3259. The images in the upper panels are oriented
  with the North up and the East left. \emph{Top left panel}: map of
  the observed surface-brightness distribution. \emph{Top
    middle panel}: map of the modeled surface-brightness
  distribution. \emph{Top right panel}: map of the residuals obtained
  by subtracting the modeled from the observed surface-brightness
  distribution. The black areas are not considered in the
  fit. \emph{Bottom left panel}: ellipse-averaged radial profile of the
  surface brightness extracted from the observed (black dots
  with gray error bars) and modeled image (green solid line) . The
  intrinsic profiles of the bulge (blue dashed line) and disk (red
  dotted line) are shown in linear and logarithmic radial scale
  (\emph{upper inset}). \emph{Bottom middle panel}: ellipse-averaged
  radial profile of the ellipticity extracted from the observed
  (black dots with gray error bars) and modeled image (green solid
  line). \emph{Bottom right panel}: ellipse-averaged radial profile of
  the position angle extracted from the observed (black dots
  with gray error bars) and modeled image (green solid line). The
  comparison between the ellipse-averaged profiles of the observed and
  model images shows the quality of the decomposition.}
\label{fig:images}
\end{center}
\end{small}
\end{figure*}}

\newcommand{\placefigurebc}{
\begin{figure}
\begin{small}
\begin{center}
\includegraphics[scale= 0.52,angle=+90]{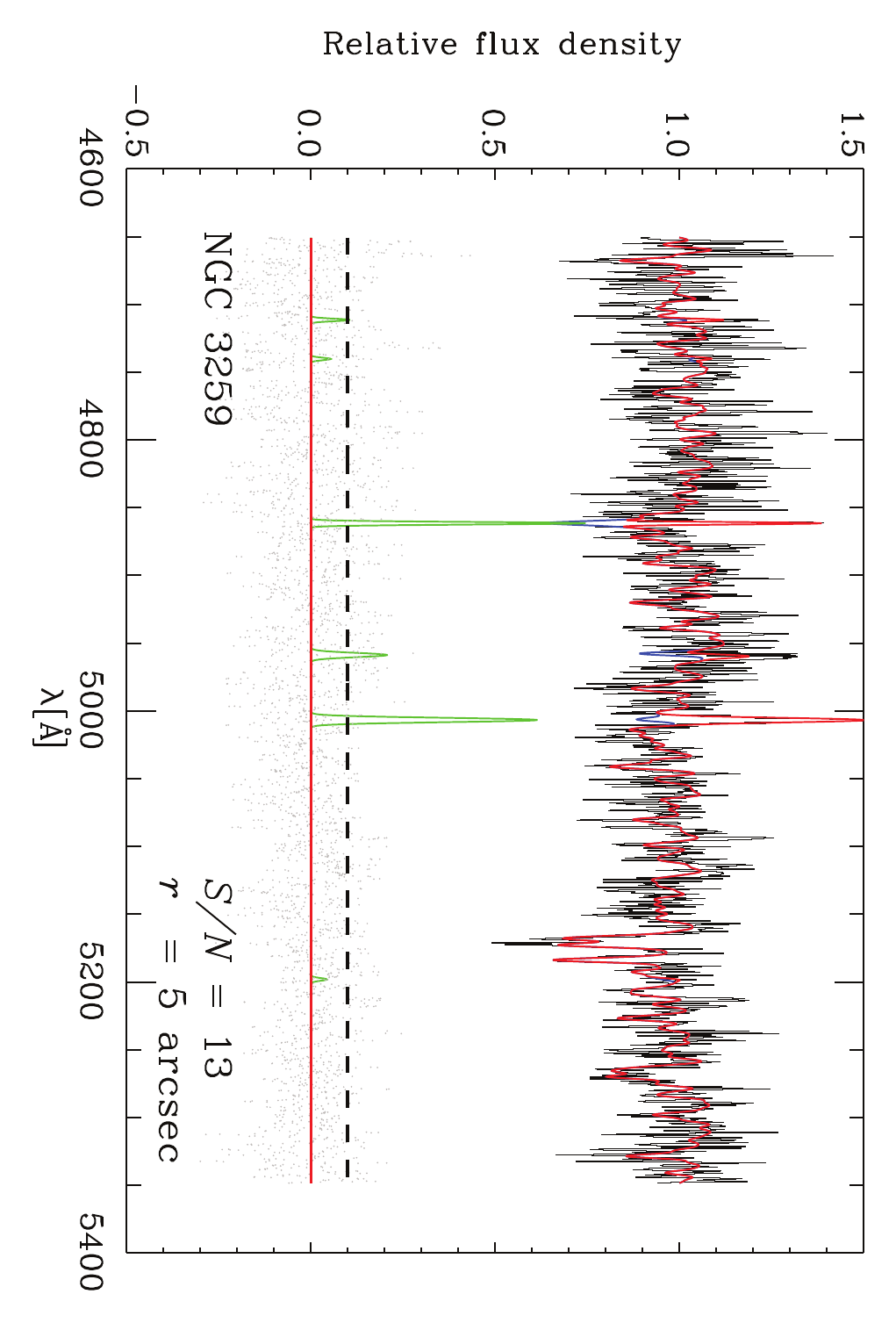}
\caption{Rest-frame T122/B\&C spectrum of NGC~3259 (black solid line)
  with the best-fitting model (red solid line) that is calculated as
  the sum of the spectra of the ionized-gas (green solid line) and
  stellar (blue solid line) components. The residuals (gray points)
  are defined as the difference between the observed and model
  spectrum. The $S/N$ is provided per resolution element and $r$ gives
  the size of the central aperture where we extracted the spectrum and
  measured \sigmas .}
\label{fig:bc}
\end{center}
\end{small}
\end{figure}}

\newcommand{\placefigurestis}{
\begin{figure}
\begin{small}
\begin{center}
\includegraphics[scale= 0.77,angle=0]{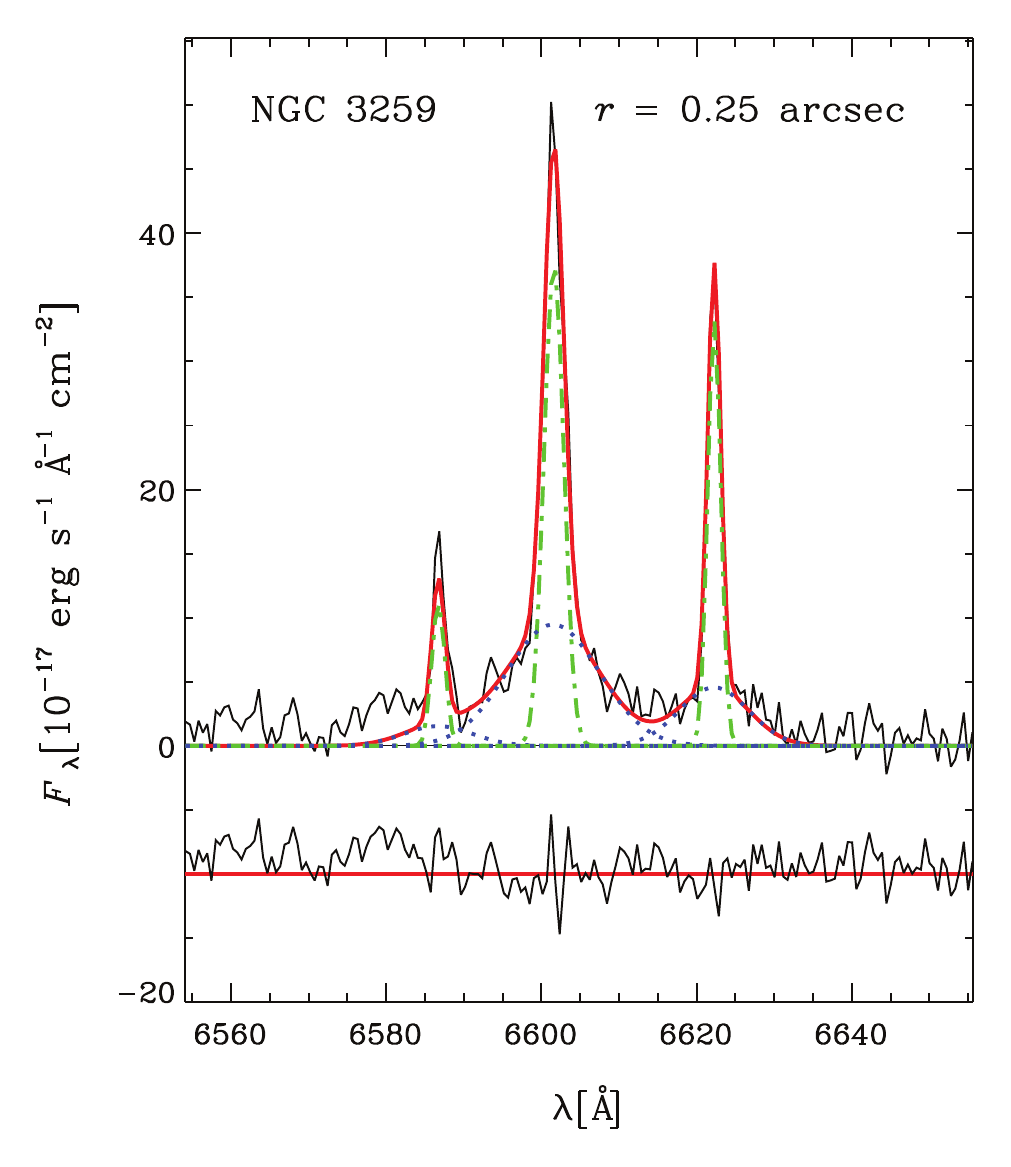}\\
\includegraphics[scale= 0.65,angle=0]{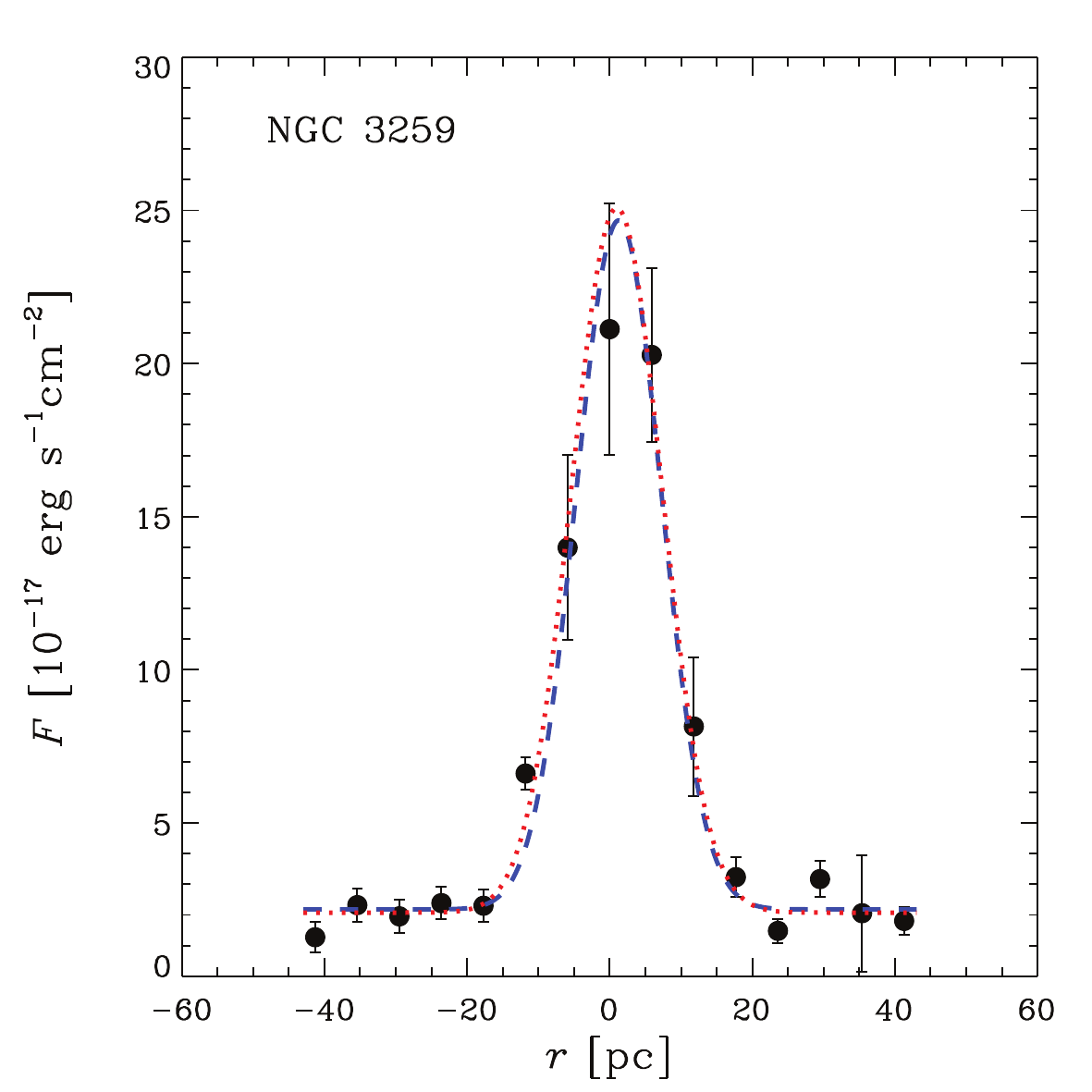}
\vspace{2mm}
\caption{\emph{Top panel}: HST/STIS spectrum of NGC~3259 (black solid
  line) with best-fitting model (red solid line) obtained adopting
  broad (blue dotted lines) and narrow (green dashed dotted lines) components
  for the \nii\ and \halpha\ lines. $r$ gives the size of the central
  aperture where we extracted the spectrum and measured \sigmag . The
  residuals are defined as the difference between the observed and
  model spectrum. They are shifted to have an arbitrary zero point for viewing
  convenience. \emph{Bottom panel}: flux radial profile of the
  \niig\ emission line measured in the HST/STIS spectrum of NGC~3259.
  The best-fitting radial profiles of intrinsic flux for a gas disk
  with $i=33^\circ$ (blue dashed line) and $81^\circ$ (red dotted
  line) are shown after being convolved with STIS PSF.}
\label{fig:stis}
\end{center}
\end{small}
\end{figure}}

\newcommand{\placefigureul}{
\begin{figure}
\begin{small}
\includegraphics[scale= 0.50,angle=0]{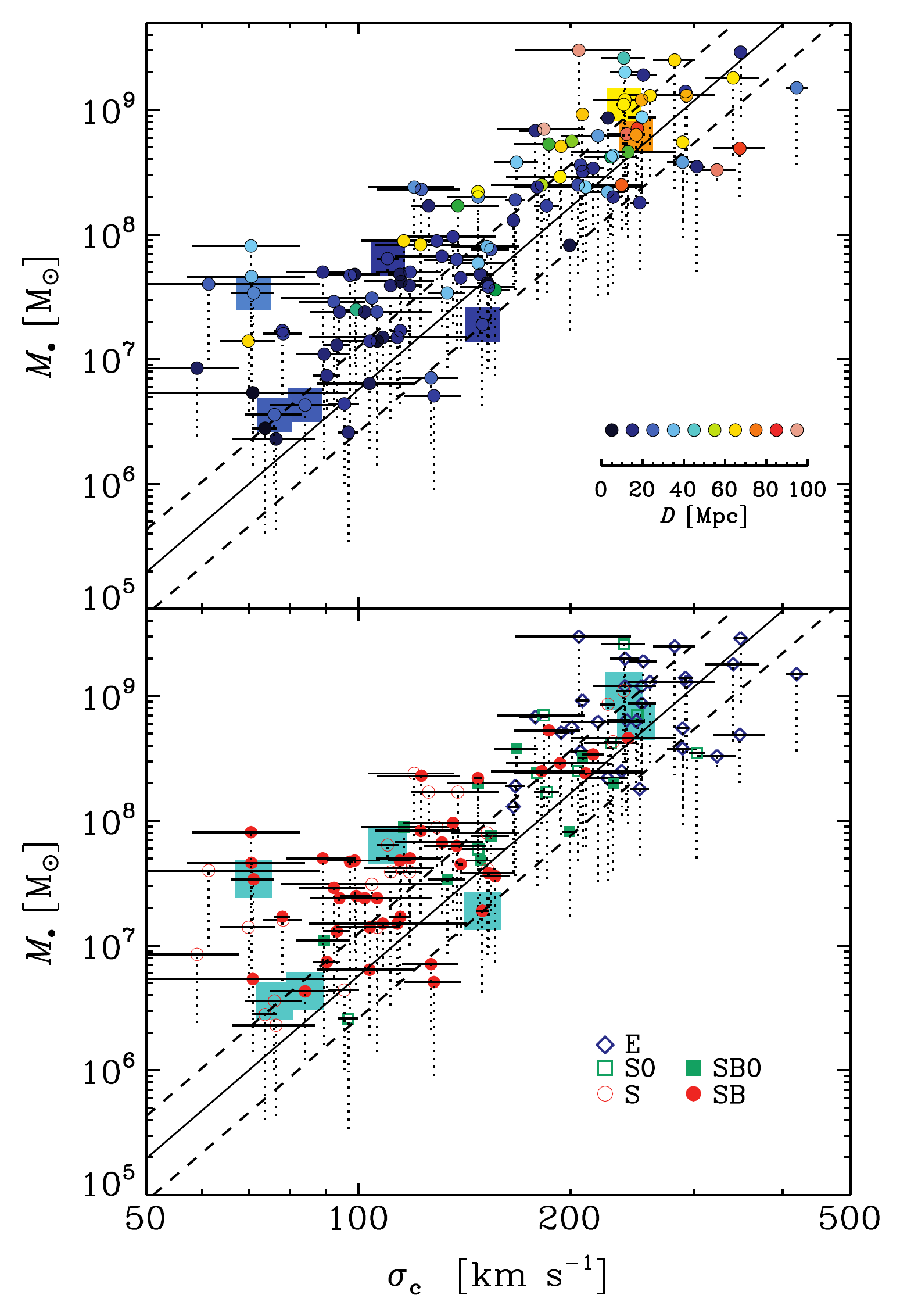}
\caption{Comparison between the stringent \mbh\ limits of our (symbols
  with large squares) and \citet{Beifiori2009, Beifiori2012} sample of galaxies 
  as a function of galaxy distance (\emph{top panel}) and morphological 
  type (\emph{bottom panel}). The upper and lower edges of the dotted 
  lines correspond to \mbh\ values estimated assuming an inclination 
  of $i = 33^\circ$ and $81^\circ$ for the gas disk, respectively. 
  In both panels the solid line is \msigmac\ relation by 
  \citet{Ferrarese2005} with the dashed lines showing the $1\sigma$ 
  (0.34 dex) scatter in \mbh .}
\label{fig:ul}
\end{small}
\end{figure}}

\newcommand{\placefigurehisto}{
\begin{figure}
\begin{small}
\includegraphics[scale= 0.4,angle=0]{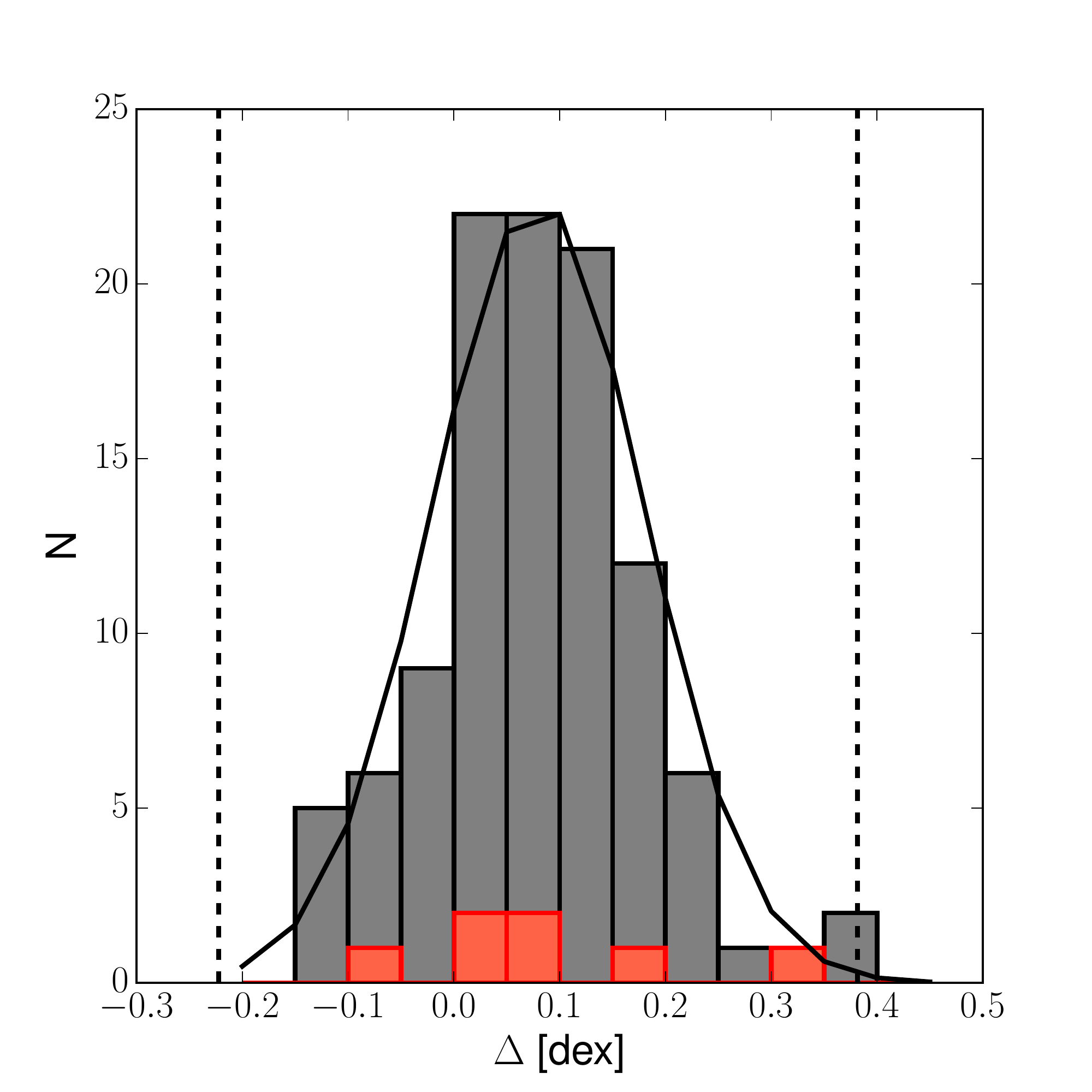}
\caption{Distribution of the distance of the (\sigmac,$\,$\mbh) values
  measured for our (red histogram) and \citet{Beifiori2009, Beifiori2012} (black
  histogram) sample of galaxies from the \msigmac\ relation by
  \citet{Ferrarese2005}. The black solid line is the Gaussian fit to
  \citet{Beifiori2009} distribution, which is centered at
  $\Delta=0.08$ dex with the black dotted lines marking its
  $\pm3\sigma$ range ($\sigma=0.10$ dex).}
\label{fig:histogram}
\end{small}
\end{figure}}

\begin{document}

\Pagespan{1}{}
\Yearpublication{2014}%
\Yearsubmission{2014}%
\Month{0}%
\Volume{999}%
\Issue{0}%
\DOI{asna.201400000}%

\title{Stringent limits on the masses of the supermassive black
  holes in seven nearby galaxies\,\thanks{Based on observations with
    the Hubble Space Telescope (HST) obtained at Space Telescope
    Science Institute (STScI), which is operated by the Association of
    Universities for Research in Astronomy (AURA), Inc., under
    National Aeronautics and Space Administration (NASA) contract
    NAS5-26555.}}

\author{I. Pagotto\inst{1}\fnmsep\thanks{Corresponding author:
        {ilaria.pagotto@phd.unipd.it}},
E.M. Corsini\inst{1,2}, E. Dalla Bont\`a\inst{1,2}, A. Beifiori\inst{3,4},
L. Costantin\inst{1}, V. Cuomo\inst{1}, L. Morelli\inst{1,2},
A. Pizzella\inst{1,2}, and M. Sarzi\inst{5}
}
\titlerunning{Stringent limits on supermassive black hole masses}
\authorrunning {I. Pagotto et al.}
\institute{Dipartimento di  Fisica e Astronomia ``G. Galilei'', 
  Universit\`a di Padova, vicolo dell'Osservatorio 3, I-35122 Padova, Italy \and
  INAF-Osservatorio Astronomico di Padova, vicolo dell'Osservatorio 2, 
      I-35122 Padova, Italy \and
  Universit\"ats-Sternwarte M\"unchen, Scheinerstra\ss e 1, 
      D-81679 M\"unchen, Germany \and
  Max-Planck-Institut f\"ur extraterrestrische Physik, Giessenbachstra\ss e, 
      D-85748 Garching bei M\"unchen, Germany \and
  Centre for Astrophysics Research, University of Hertfordshire, 
      College Lane, Hatfield AL10 9AB, UK
}

\received{XXXX}
\accepted{XXXX}
\publonline{XXXX}


\abstract{We present new stringent limits on the mass \mbh\ of
  the central supermassive black hole for a sample of 7 nearby
  galaxies.  Our \mbh\ estimates are based on the dynamical modeling
  of the central width of the nebular emission lines measured over
  subarcsecond apertures with the Hubble Space Telescope. The central
  stellar velocity dispersion \sigmac\ of the sample galaxies is
  derived from new long-slit spectra from ground-based
  observations and the bulge effective radius is obtained from a
  two-dimensional photometric decomposition of the $i$-band images
  from the Sloan Digital Sky Survey. The derived stringent \mbh\ limits
  run parallel and above the \msigmac\ relation with no systematic
  trend depending on the galaxy distance or morphology. This gives
  further support to previous findings suggesting that the nuclear
  gravitational potential is remarkably well traced by the width of
  the nebular lines when the gas is centrally peaked. With our 
  investigation, the number of galaxies
  with stringent \mbh\ limits obtained from nebular-line width increases
  to 114 and can be used for studying the scaling relations between
  \mbh\ and properties of their host galaxies.}

\maketitle

\section{Introduction}
\label{sec:introduction}

Over nearly three decades of measurements for the mass \mbh\ of central 
supermassive black holes (SBHs) have led to the conclusion that such objects 
should be nearly always present at the center of elliptical galaxies and bulges
of disk galaxies \citep[see][for a review]{Kormendy2013}. 
Furthermore, the finding that \mbh\ correlates with several properties of 
their host galaxies, and in particular with the velocity dispersion \sigmas\ 
of their spheroidal component (\citealt{Gebhardt2000}; \citealt{Ferrarese2000}), 
suggests that somehow SBHs and spheroids grew together 
(see \citealt{Saglia2016}). These mutual relationships between the black 
holes and their host galaxies could come from feedback mechanisms 
(see \citealt{Silk1998}; \citealt{Fabian1999}). Large \mbh\ samples across 
different morphological types are needed to fully understand the underlying 
process behind the \msigmas\ relation, as clues may be present not only in 
the slope of this relation but also in its scatter, the behavior of outliers, 
and secondary trends related for instance to galaxy morphology 
(\citealt{Beifiori2012}; \citealt{McConnell2013}; 
\citealt{Shankar2016}).

For this purpose \citet{Beifiori2009, Beifiori2012} used archival Hubble 
Space Telescope (HST) spectroscopic data obtained with the Space Telescope 
Imaging Spectrograph (STIS) to estimate stringent limits on \mbh\ for 107 
galaxies of various Hubble types, following the approach of \citet{Sarzi2002} 
to model the velocity dispersion of ionized-gas emission observed at sub-arcsecond 
scales. The sample of \citet{Beifiori2009} included an additional 21 objects 
with STIS data, but those were excluded since they missed a ground-based 
\sigmas\ measurement and thus could not be placed on the \msigmas\ relation. 
In this paper, we aim at increasing the sample of galaxies with stringent 
$M_{\bullet}$ limits, by selecting 7 Northern galaxies ($\rm
Dec\,(J2000.0)\,>\,-10^\circ$) with detected emission lines in STIS 
spectra from the objects excluded by \citet{Beifiori2009} and observing 
them with the Asiago Astrophysical Observatory in order to derive their 
central \sigmas. 

\placetablegalsample

The paper is organized as follows. We derive the bulge effective
radius $r_{\rm e}$ from the analysis of broad-band imaging in
Section~\ref{sec:star_photometry}. We measure 
the central stellar velocity dispersion from ground-based spectroscopy
and apply the aperture correction to $r_{\rm e}/8$ in
Section~\ref{sec:star_kinematics}. We obtain the distribution and
central velocity dispersion of the ionized gas from HST spectroscopy
in Section~\ref{sec:gas_kinematics}. We estimate the stringent \mbh\ limit
from gas dynamics in Section~\ref{sec:dynamics}.  Finally, we discuss
our results in the framework of the \msigmas\ relation in
Section~\ref{sec:conclusion}. In this work we adopt 
$H_{0} = \rm 75~km~s^{-1}~Mpc^{-1}$, 
$\Omega_{\rm M} = 0.3$, and $\Omega_{\Lambda}= 0.7$ 
as cosmological parameters.

\section{Surface-brightness distribution}
\label{sec:star_photometry}

\subsection{Sloan Digital Sky Survey imaging}

We retrieved the flux-calibrated $i$-band images of the sample
galaxies from the Data Release 12 of the Sloan Digital Sky Survey
\citep[SDSS-DR12,][]{Alam2015}.  

We measured the sky level to be subtracted from the image of each
sample galaxy, as done in \citet{Morelli2016}. We masked the stars,
galaxies, and spurious sources in the galaxy neighborhoods and
measured its surface brightness radial profile with the ELLIPSE task
in IRAF{\footnote{Image Reduction and Analysis Facility is distributed
    by the National Optical Astronomy Observatory (NOAO), which is
    operated by the AURA, Inc., under cooperative agreement with the
    National Science Foundation.}. First, we fitted the galaxy
  isophotes with ellipses having the center, ellipticity, and position
  angle free to vary. Then, we repeated the isophotal fit fixing
  the center we previously obtained for the inner ellipses and the
  ellipticity and position angle of the outer ones. We calculated the
  sky level by averaging the surface brightness measured at large
  radii, where there is no light contribution from the galaxy. We used
  the IRAF task IMEXAMINE to measure the standard deviation of the
  background in the sky-subtracted images and to fit the stars of the
  field of view with a circular Moffat profile \citep{Moffat1969},
  which we adopted to model the point spread function (PSF).
Finally, we trimmed the sky-subtracted images to reduce the computing
time required to perform a reliable photometric decomposition and we ran 
ELLIPSE on the trimmed images to derive the radial profiles
of surface brightness, ellipticity, and position angle. They were used
as an input to get the starting guesses of the galaxy structural
parameters for the two-dimensional photometric decomposition.

\subsection{Photometric decomposition}

\placefiguredecone

To measure the effective radius of the bulge, we performed the 
two-dimensional photometric decomposition of the SDSS images of the sample galaxies
by using the Galaxy Surface Photometry 2-Dimensional Decomposition
algorithm \citep[GASP2D,][]{MendezAbreu2008, MendezAbreu2014}.

GASP2D performs a two-dimensional parametric photometric decomposition assuming that
the observed surface brightness of the galaxy in each image pixel is
expressed as the sum of analytical functions describing the light
contribution of the structural components. We modeled (1)
the surface brightness of the bulge with a S\'ersic law
\citep{Sersic1968}
\begin{equation}
I_{\rm bulge}(r) = I_{\rm e} 10^{-b_n[(r/r_{\rm e})^{1/n} -1 ]} ,
\end{equation}
where $r_{\rm e}$ is the effective radius, $I_{\rm e}$ is the
surface brightness at $r_{\rm e}$, $n$ is the shape parameter of the
surface brightness profile, and $b_n = 0.868n - 0.142$
\citep{Caon1993} is a normalization coefficient;
(2) the surface brightness of the disk either with a
single exponential law \citep{Freeman1970}
\begin{equation}
I_{\rm disk}(r) = I_{\rm 0} e^{-r/h} ,
\end{equation}
where $I_{\rm 0}$ is the central surface brightness and $h$ is the
scalelength, or with a double-exponential law \citep{vanderKruit1979}
\begin{equation}
I_{\rm disk}(r) = I_0 e^{-r_{\rm break} (h_{\rm out} - h)/(h_{\rm out} h)} e^{-r/h} ,
\end{equation}
where $I_{\rm 0}$ is the central surface brightness, $r_{\rm break}$
is the break radius at which the slope change occurs,
$h$ and $h_{\rm out}$ are the scalelengths of the inner and
outer exponential profiles, respectively;
(3) the surface brightness of the bar with a Ferrers law
(\citealt{Ferrers1877}; \citealt{Aguerri2009})
\begin{equation}
I_{\rm bar}(r) =
\begin{cases}
I_{\rm 0,bar} \biggl[1 - \biggl(r/a_{\rm bar}\biggl)^2 \biggr]^{2.5} & 
\text{if $r \leq a_{\rm bar}$} \\
0 & \text{if $r > a_{\rm bar}$} ,
\end{cases}
\end{equation}
where $I_{\rm 0,bar}$ is the central surface brightness and $a_{\rm
  bar}$ is the bar length.
We assumed the isophotes of the bulge, disk, and bar to be elliptical,
centered onto the galaxy center, and with constant position angle 
${\it PA}_{\rm bulge}$, ${\it PA}_{\rm disk}$, and ${\it PA}_{\rm bar}$ and
constant axial ratio $q_{\rm bulge}$, $q_{\rm disk}$, and $q_{\rm
  bar}$, respectively. We did not considered other components, such as
rings, lenses, ovals, or spiral arms. 

\placetablegasptwod

GASP2D returns the best-fitting values of the structural parameters
of the bulge ($I_{\rm e}$, $r_{\rm e}$, $n$, ${\it PA}_{\rm bulge}$, $q_{\rm
bulge}$), disk ($I_{\rm 0} $, $h$, $h_{\rm out}$, $r_{\rm b}$, 
${\it PA}_{\rm disk}$, $q_{\rm disk}$), and bar ($I_{\rm 0,bar}$, $a_{\rm bar}$,
${\it PA}_{\rm bar}$, $q_{\rm bar}$) with a $\chi^2$ minimization by
weighting the surface brightness of the image pixels according to the
variance of the total observed photon counts due to the contribution
of both galaxy and sky. It accounts as well for photon noise, CCD gain
and read-out noise, and image PSF. We derived the errors on the
structural parameters by analyzing a sample of mock galaxies generated
with Monte Carlo simulations, as done by \citet{Costantin2017}.

We successfully performed the photometric decomposition of all the
sample galaxies, except for NGC~2654, NGC~4343, and NGC~5635.
For the barred galaxy NGC~3049, we adopted the double-exponential disk
to fit the slope change measured in the surface brightness distribution
at about 50 arcsec.
We fitted NGC~3259 with a S\'ersic bulge and an exponential disk,
although it is classified as a weakly barred galaxy. Indeed, an
accurate inspection of the image shows that what it looks like a bar
is actually an artifact resulting from the tightly wound spiral arms.
NGC~5141 is classified as a lenticular galaxy, but we successfully
fitted its surface brightness with a single S\'ersic component.
The best-fitting model was chosen by applying the Bayesian information 
criterion \citep{Schwarz1978} as done in 
\citet{MendezAbreu2017a, MendezAbreu2017b} to discriminate between 
ellipticals and lenticulars. Therefore, the best-fitting value 
of $r_{\rm e}$ refers to the entire galaxy.
The prominent spiral arms of NGC~5713 produce the abrupt changes in
surface brightness, ellipticity, and position angles measured at about
30 arcsec. We included the bar component in the fit in order to not 
overestimate the bulge contribution.
We show in Fig.~\ref{fig:images} the photometric decomposition of
NGC~3259 as an example and report the results for the sample
galaxies we analyzed in Table~\ref{tab:gasp2d}. We present 
in Fig.~\ref{fig:appimages} the photometric decompositions
of the remaining sample galaxies.

The GASP2D photometric decomposition of the $i$-band SDSS image of
NGC~5635 was performed by \citet{MendezAbreu2017a} and we took for the
bulge their best-fitting value of $r_{\rm e}=8.0\pm0.9$ arcsec.

NGC~2654 and NGC~4343 are two nearly edge-on galaxies. This prevented
us from performing the photometric decomposition with GASP2D, which is
best designed for galaxies with low-to-intermediate inclination
\citep[see][for a discussion]{MendezAbreu2008}. We adopted the bulge
$r_{\rm e}$ obtained by \citet[][NGC~2654: $r_{\rm e}=3.9$ arcsec,
NGC~4343: $r_{\rm e} = 18.7$ arcsec]{Salo2015} from the analysis of the
infrared images of the Spitzer Survey of Stellar Structure in Galaxies
\citep[S$^4$G,][]{Sheth2010}.

\section{Stellar kinematics}
\label{sec:star_kinematics}

\subsection{Ground-based spectroscopy}

\placefigurebc

We performed the spectroscopic observations of the sample galaxies
from 2016 January 4 to May 5 with the 1.22-m Galileo telescope (T122)
at the Asiago Astrophysical Observatory (Italy). We used the Boller \&
Chivens spectrograph (B\&C) and the grating with 1200 grooves
mm$^{-1}$ in combination with a 2 arcsec $\times$ 7.75 arcmin slit and the
Andor iDus DU440 CCD composed by $2048 \times 512$ pixels of 26 $\mu$m
$\times$ 26 $\mu$m each. The CCD gain and readout noise rms were 0.97
$e^{-}\,$ADU$^{-1}$ and 3.4 $e^{-}$, respectively. The spectra were
characterized by a wavelength range between about $4500\,$--$\,5700$ $\AA$ with a
reciprocal dispersion of 0.60 \apix\ and a spatial scale of 1.0 arcsec 
pixel$^{-1}$.
We estimated the instrumental resolution as the mean of the Gaussian
FWHMs measured for the unblended emission lines of a
wavelength-calibrated comparison spectrum. We found ${\it FWHM}_{\rm inst} =
1.578\pm0.002$ $\AA$ corresponding to $\sigma_{\rm inst} \simeq 40$ \kms\ at
5100 $\AA$. During the observing nights, we measured a seeing
 ${\it FWHM}_{\rm PSF}=3\,$--$\,5$ arcsec by fitting the guiding star with a circular
Gaussian function.

\placetableasiago

We observed all the sample galaxies by centering their nucleus into
the slit which we aligned along the galaxy major axis according to the
position angle tabulated in Table~\ref{tab:asiago}. We splitted the
total integration time into single exposures of 1800 s each for a better
rejection of cosmic ray events. At the beginning of each exposure, we moved
the slit of 10 arcsec along the galaxy major axis in order to avoid that
bad pixels occurred in the same CCD position. During each night, 
we observed some spectrophotometric standard stars from the list of
\citet{Hamuy1992, Hamuy1994} to successfully perform the flux calibration. 
Finally, we obtained a
comparison spectrum of the available HeFeAr arc lamp before and after
each object exposure to ensure an accurate wavelength calibration. 
The total exposure time of the T122 spectra is
reported in Table \ref{tab:asiago}.

We reduced the T122 spectra using standard tasks in IRAF, as done in
\citet{Corsini2017}. The reduction steps included the subtraction of
bias, correction for internal and sky flat-field, trimming of the
spectra, removal of bad pixels and cosmic rays, correction for CCD
misalignment, subtraction of the sky contribution, wavelength and flux
calibration, alignment, and combination of the spectra obtained for the
same galaxy. The combined T122 spectrum of NGC~3259 is given in
Fig.~\ref{fig:bc}. The T122 spectra of the remaining sample galaxies  
are presented in Fig.~\ref{fig:appT122}.

\subsection{Absorption-line fitting}

To measure the central stellar velocity dispersion, we analyzed the
absorption lines in the combined T122 spectra of the sample galaxies
with the Penalized Pixel Fitting \citep[pPXF,][]{Cappellari2004} and
Gas and Absorption Line Fitting \citep[GANDALF,][]{Sarzi2006}
algorithms.

For each galaxy, we rebinned the combined spectrum along the
dispersion direction to a logarithmic scale and we deredshifted 
it to rest frame. Then, we averaged the rebinned spectrum along the 
spatial direction either to cover a nearly square aperture 
(Table~\ref{tab:asiago}) or to have a signal-to-noise
ratio $S/N \geq 10$ per resolution element.
We convolved a linear combination of stellar spectra from the ELODIE
library at medium resolution \citep[$\sigma_{\rm inst} = 13$
  \kms,][]{Prugniel2001} with a Gauss-Hermite line-of-sight velocity
distribution \citep[LOSVD,][]{Gerhard1993, vanderMarel1993} to fit the
summed spectrum by a $\chi^2$ minimization in pixel space. We selected 
229 stellar spectra to fully cover the ELODIE
parameter space of the effective temperature, surface gravity, and
metallicity. They were broadened to match the T122 instrumental
resolution, logarithmically rebinned, and cropped along the 
dispersion direction in order to match the wavelength range of each 
galaxy spectrum. 

In addition, we simultaneously fitted all the ionized-gas 
emission lines in the covered wavelength range. We also 
added a fourth-order multiplicative Legendre polynomial
to correct for reddening and large-scale residuals of flat-fielding. 
We excluded from the fitting procedure the wavelength ranges with 
a spurious signal coming from imperfect
subtraction of cosmic rays and sky emission lines.

First, we obtained the best-fitting values of the LOS velocity
$v_\star$, velocity dispersion \sigmas , and Gauss-Hermite
coefficients $h_3$ and $h_4$ of the stellar component. The fitting
procedure returned the value of \sigmas\ corrected for instrumental
resolution. After checking that $h_3 = h_4 = 0$ within the errors, we
fitted again the galaxy spectra adopting a Gaussian LOSVD to measure
\sigmas\ (Table~\ref{tab:asiago}). We assumed its error to be the
formal error of the best fit after evaluating $\chi^2$ to achieve
$\chi^2=N_{\rm dof}=N_{\rm d}-N_{\rm fp}$, where $N_{\rm dof}$,
$N_{\rm d}$, and $N_{\rm fp}$ are the number of the degrees of
freedom, data points, and fitting parameters, respectively
\citep{Press1992}. The best-fitting model of the T122 spectrum of
NGC~3259 is displayed in Fig.~\ref{fig:bc}.
Finally, for full consistency with \citet{Beifiori2009, Beifiori2012} 
we applied the aperture correction of \citet{Jorgensen1995}
to obtain the stellar velocity dispersion \sigmac , that would have been
measured within a circular aperture of radius $r_{\rm e}/8$
(Table~\ref{tab:sample}).

\section{Ionized-gas distribution and kinematics}
\label{sec:gas_kinematics}

\subsection{Hubble Space Telescope spectroscopy}

\placetablehst

From the Hubble Data Archive, we retrieved the STIS spectra of the
sample galaxies obtained with the G750M grating through either the
$0.1 \times 52$ arcsec$^{2}$ or the $0.2 \times 52$ arcsec$^{2}$ slit 
placed across the galaxy nucleus at position angle close to the 
galaxy major axis. The detector was a SITe CCD with $ 1024\times1024$ 
pixel of $21\times21$ $\mu$m$^2$. The spectra covered a
wavelength range of either $6480\,$--$\,7060$ or $6300\,$--$\,6870$ $\AA$, depending on
the tilt angle of the grating. The reciprocal dispersion was 0.554 and
1.108 \apix\ for the 1-pixel and 2-pixel binning read-out mode along
the dispersion direction, respectively. This setup yielded an
instrumental FWHM of 0.87 $\AA$ ($\sigma_{\rm inst} = 17$ \kms) in the
case of a 0.1 arcsec-wide slit and 1.6 $\AA$ ($\sigma_{\rm inst} = 32$
\kms) for the 0.2 arcsec-wide slit (\citealt{Sarzi2002}; \citealt{Beifiori2009}). 
The spatial scale was 0.0507 arcsec and 0.101 arcsec pixel$^{-1}$ for 
the 1-pixel and 2-pixel binning read-out mode along the spatial direction, 
respectively. The HST proposal number, slit width and position angle, 
pixel binning, wavelength range, and total exposure times of the HST 
spectra of the sample galaxies are reported in Table~\ref{tab:hst}.

We reduced the HST spectra using IRAF tasks, as done in
\citet{Beifiori2009}. The reduction steps included the subtraction of
the overscan, bias and dark contributions, correction for internal
flat-field, trimming of the spectra, removal of bad pixels and cosmic
rays, wavelength and flux calibration, correction for geometrical
distortion, alignment and combination of the spectra obtained for the
same galaxy. The combined HST spectrum of NGC~3259 is shown in
Fig.~\ref{fig:stis}. We present in Fig.~\ref{fig:appstis} 
the HST spectra of the remaining sample galaxies.

\subsection{Emission-line fitting}

\placefigurestis

To measure the distribution and kinematics of the ionized gas, we
analyzed the \nii , \halpha, and \sii\ emission lines in the combined
HST spectra of the sample galaxies, following the prescriptions by
\citet{Beifiori2009} and using their IDL algorithm based on MPFIT 
package \citep{Markwardt2009}.

We fitted the stellar continuum with a low-order polynomial and the
narrow and broad components of the observed emission lines with a sum
of Gaussian functions. For all the objects the ionized-gas emission 
was always much stronger compared to the stellar continuum. 
The best-fitting parameters and their errors
were derived by a $\chi^2$ minimization in pixel space and evaluating
$\chi^2=N_{\rm dof}$. We adopted both a narrow and broad component for
the emission lines of NGC~3259 and NGC~5141, two
narrow components for NGC~3049, and a single narrow component for all
the other sample galaxies, including NGC~5635 for which the presence
of two distinct components was not clear. For NGC~2654 and NGC~3049 
the \halpha\ line was not accurately fitted due 
to the low $S/N$ and presence of an asymmetric broad component, respectively.
This does not influence our results since we considered only the 
\niig\ line for the dynamical modeling.

We focused on the \niig\ line because it was always the brightest
nebular line in our spectra. The nebular lines actually probe the
nuclear kinematics better than the \halpha\ line, which could be
affected by both the absorption from the stellar component and
emission from circumnuclear starforming regions
\citep[e.g.,][]{Coccato2006}.

We measured the radial profile of the \niig\ flux along the spatial
direction to constrain the distribution of the ionized gas. We assumed 
the gas to be distributed into a infinitesimally thin disk centered on the
galaxy nucleus with an intrinsically Gaussian flux profile. We derived
the intrinsic flux profile for two inclinations of the gaseous disk
($i=33^\circ, 81^\circ$) by a $\chi^2$ minimization to match the
observed flux, while accounting for the STIS PSF, which we generated
with the TINY TIM package by \citet{Krist2011}. The result for
NGC~3259 is plotted in Fig.~\ref{fig:stis}.
We measured the \niig\ line width within a nearly square aperture
centered on the continuum peak to estimate the central gas velocity
dispersion \sigmag . In the case of the 1-pixel spatial binning, we
considered an aperture of 0.15 arcsec (3 pixels) or 0.25 arcsec (5
pixels) along the spatial direction when the spectrum was obtained
with a 0.1 arcsec or a 0.2 arcsec-wide slit, respectively. In the case of
a 2-pixel spatial binning, we took an aperture of 0.30 arcsec (3
pixels). The aperture sizes are listed in Table~\ref{tab:hst}.  We
corrected the measured line width for instrumental resolution to
obtain \sigmag\ (Table \ref{tab:hst}).

\section{Dynamical modeling}
\label{sec:dynamics}

To derive stringent \mbh\ limits for our sample galaxies 
we followed the approach of \citet{Sarzi2002}. This is based on 
the assumption that the nuclear gravitational potential is traced by the
line width of the nebular emission originated from ionized gas in
Keplerian rotation around the SBH. We considered the gas moving onto
coplanar circular orbits in a infinitesimally thin disk with the
intrinsic distribution we derived from the measured flux radial
profile.
The kinematics was measured along only one direction across the galaxy
nucleus, which does not provide constraints on the orientation of the
gas disk. We disregarded the effect on the unknown position angle of
the gas disk, since we extracted our spectrum in a nearly square
aperture. Thus, we could assume that the STIS slit was placed along
the major axis of the gas disk and we estimated \mbh\ at two
inclinations of $33^{\circ}$ and $81^{\circ}$, which bracket the
$68\%$ of randomly inclined disks \citep[see][]{Sarzi2002}. 
Our choice of parametrization for the intrinsic flux of the gas is
conservative, a cuspier function would
lead to a more concentrated gas distribution and, therefore, to a
smaller value of \mbh\ \citep[see][]{Sarzi2002}. We also expect to find a
smaller value of \mbh\ when taking into account the contributions of
the stellar potential and non-gravitational forces (due for instance to 
the activity of the central SBH). These last factors imply that our 
\mbh\ estimates based on the assumption of a rotating gaseous 
disk should be strictly speaking regarded only as upper limits, although 
in practice it is unlikely that the gas motions are entirely driven by 
non-gravitational forces. In particular, the finding that the narrow-line 
ionized-gas emission is always rather concentrated around the 
nucleus contrasts with the idea that such gas 
would not respond to the central gravitational pull of the SBH. In fact, 
that non-gravitational forces are generally unimportant is also supported 
by the finding that the \mbh\ estimates derived from the central ionized-gas 
flux profile and line-width always agree with actual \mbh\ measurements based 
on resolved stellar and ionized-gas kinematics \citep{Beifiori2012}. 
Nonetheless, in some instances there are indications that \mbh\ estimates based on 
our method could be biased either due to the presence of non-gravitational 
forces or to the stellar potential \citep{Beifiori2009}.

\section{Discussion and conclusions}
\label{sec:conclusion}

\placefigureul

In Fig.~\ref{fig:ul}, we compare our stringent \mbh\ limits to those of
\citet{Beifiori2009, Beifiori2012} who considered the \msigmac\ relation by
\citet{Ferrarese2005}.

The \mbh\ limits of NGC~2654, NGC~3049, NGC~3259, NGC~4343,
NGC~5141, and NGC~5635 are within the $3\sigma$ scatter of the
\msigmac\ relation. The \mbh\ of NGC~5635 is displaced towards higher
values because we probably overestimated the \sigmag\ value from the
HST/STIS spectrum, where it was not possible to distinguish the broad
and narrow components of the \niig\ line.
The \mbh\ value for NGC~5713 exceeds more than
three times the scatter of the \msigmac\ relation. For this galaxy
the contribution of the stellar mass that we disregarded in our
analysis could play a significant role, as already pointed out by
\citet{Beifiori2009} for a number of other galaxies with a
stringent \mbh\ limit in the lower end of the \msigmac\ relation. On the
top of this, the center of NGC~5713 is poorly constrained because 
its nuclear surface brightness distribution is characterized by 
several bright knots rather than a well defined peak 
\citep[see also][]{Scarlata2004}. 

Our galaxies cover different morphological types (1 elliptical, 3
unbarred and 3 barred spirals according to the photometric
decomposition) and span a wide range of central stellar velocity
dispersion ($71 < \sigma_{\rm c} < 248$ \kms ) and
\mbh\ ($3.6\cdot10^6 < M_\bullet < 1.1\cdot10^9$ \msun\ for $i = 33^{\circ}$
and $1.0\cdot10^6 < M_\bullet < 2.9\cdot10^8$ \msun\ for $i =
81^{\circ}$). The stringent \mbh\ limits we measured are fully 
consistent with those by \citet{Beifiori2009, Beifiori2012}. This is
confirmed by the distribution of the distances of the two
(\sigmac,$\,$\mbh) datasets from the \msigmac\ relation by
\citet{Ferrarese2005}, which we show in Fig.~\ref{fig:histogram} for
$i = 33^{\circ}$. On average, our \mbh\ run parallel and above the
\msigmac\ relation with no systematic trend depending on the galaxy
distance or presence of the bar.

\placefigurehisto

With our investigation, the number of galaxies with stringent \mbh\ limits
obtained from nebular-line width increases to 114 and can be used for
studying the scaling relations between \mbh\ and properties of their
host galaxies. Most of our \mbh\ limits actually populate the
low-$\sigma$ end of the \msigmac\ relation. They could be used to
prove the claim by \citet{Beifiori2009} that the contribution of the
stellar component to the gravitational potential is particularly
significant in the low-$\sigma$ regime and biases the measured
\mbh\ towards exceedingly large values. This will be investigated in 
a forthcoming paper (Pagotto et al. in prep.) to prove 
previous findings suggesting that the nuclear gravitational
potential is remarkably well traced by the width of the nebular lines
when observed at sub-arcsecond scales.

%
%

\acknowledgements
We acknowledge the anonymous referee for valuable comments 
that led to an improved presentation.
We are grateful to Heikki Salo for valuable
discussion on the photometric decomposition of NGC~4343. EMC, EDB, LM,
and AP are supported by Padua University through grants 60A02-5857/13,
60A02-5833/14, 60A02-4434/15, CPDA133894, and BIRD164402/16. IP
acknowledges the Max-Planck-Institut f{\"u}r extraterrestrische Physik
for the hospitality while this paper was in progress. Part of the data
used in this paper were acquired through the Sloan Digital Sky Survey
Archive (http://www.sdss.org/).  This research also made use of
the HyperLeda Database (http://leda.univ-lyon1.fr/) and NASA/IPAC
Extragalactic Database (http://ned.ipac.caltech.edu/).

\begin{small}

\end{small}

\appendix

\section{Appendix}
\label{Sec:App}
Fig.~\ref{fig:appimages}, Fig.~\ref{fig:appT122}, 
and Fig.~\ref{fig:appstis} show the two-dimensional 
photometric decomposition of the $i$-band SDSS images, the 
rest-frame T122/B\&C spectra, and HST/STIS spectra of the 
sample galaxies, except for NGC~3259, respectively.

\begin{figure*}
\begin{small}
\begin{center}
\includegraphics[scale= 0.54,angle=-90]{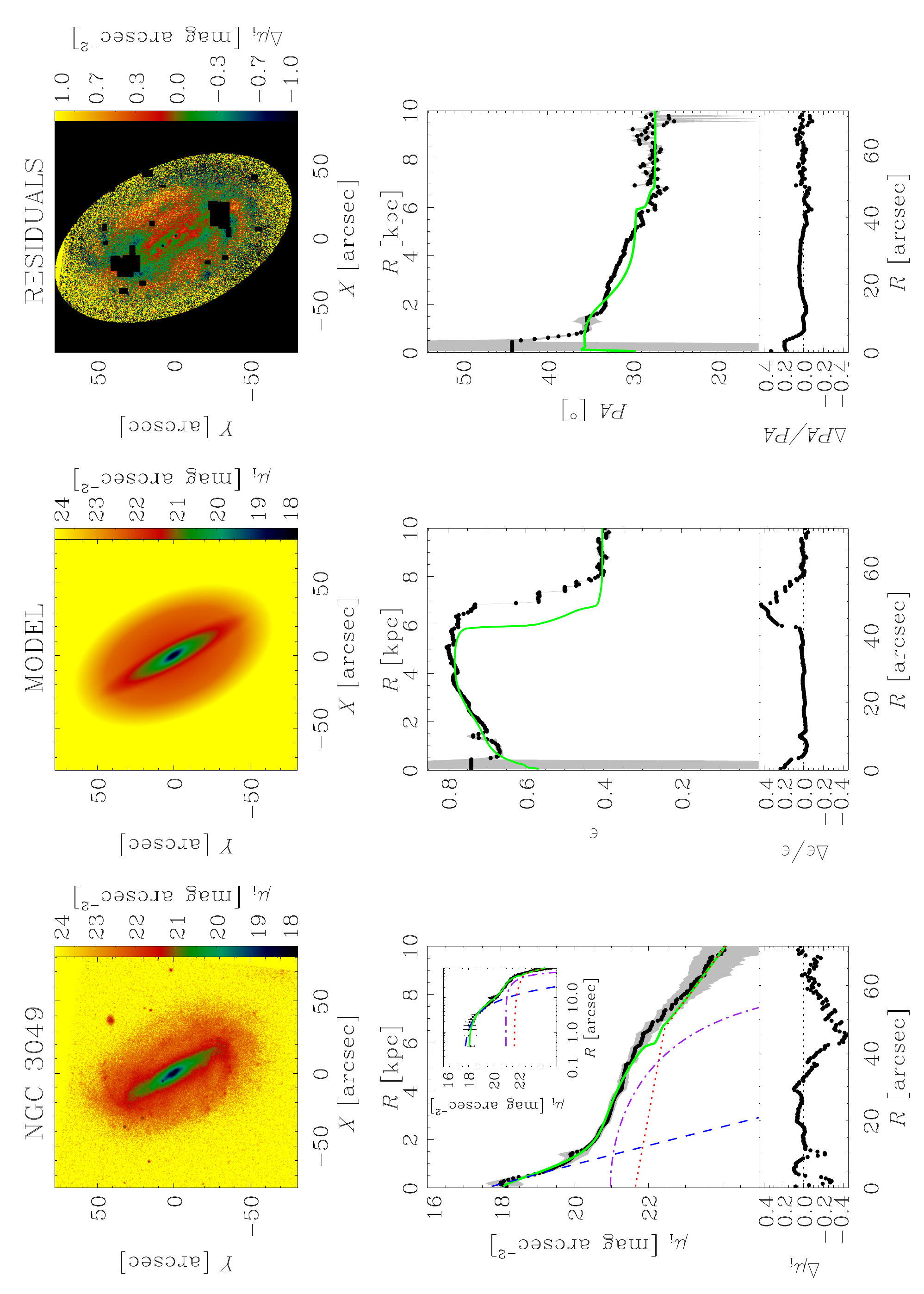}
\caption{As in Fig.~\ref{fig:images}, but for the remaining sample galaxies.}
\label{fig:appimages}
\end{center}
\end{small}
\end{figure*}

\begin{figure*}
\setcounter{figure}{0}
\begin{small}
\begin{center}
\includegraphics[scale= 0.54,angle=-90]{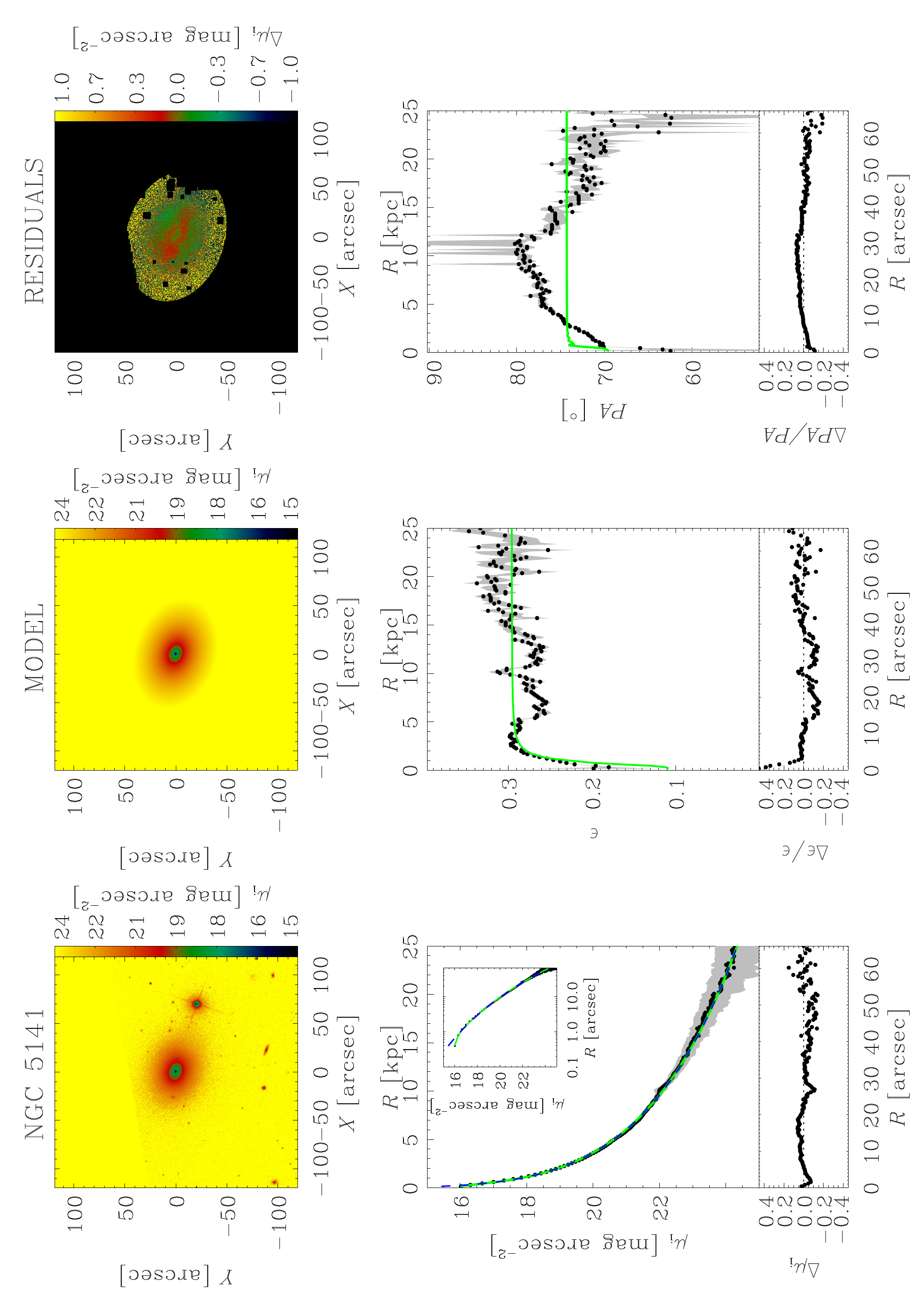}
\caption{Continued.}
\label{fig:appimages}
\end{center}
\end{small}
\end{figure*}

\begin{figure*}
\setcounter{figure}{0}
\begin{small}
\begin{center}
\includegraphics[scale= 0.54,angle=-90]{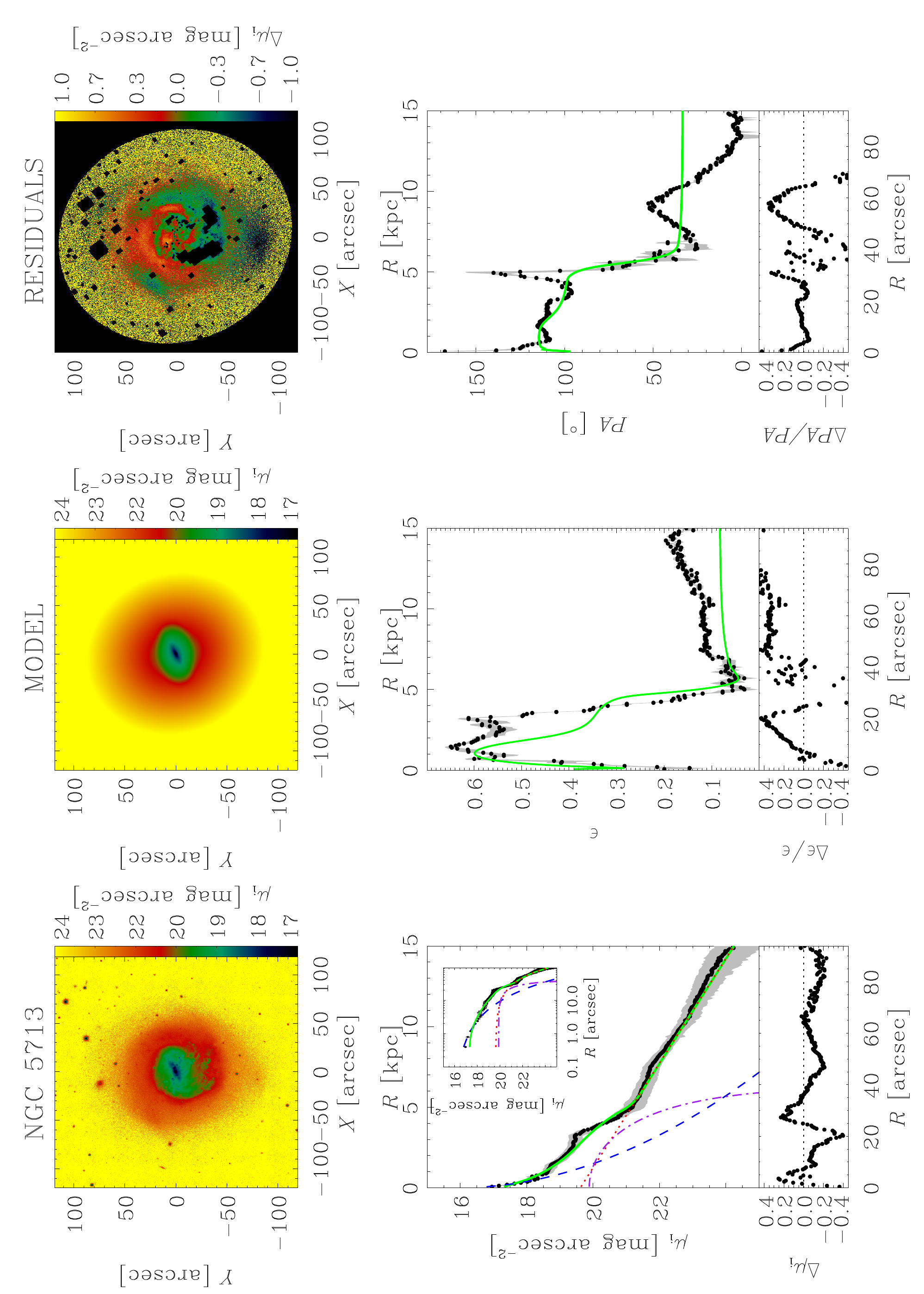}
\caption{Continued.}
\label{fig:appimages}
\end{center}
\end{small}
\end{figure*}

\begin{figure*}
\begin{small}
\begin{center}
\includegraphics[scale= 0.75,angle=0]{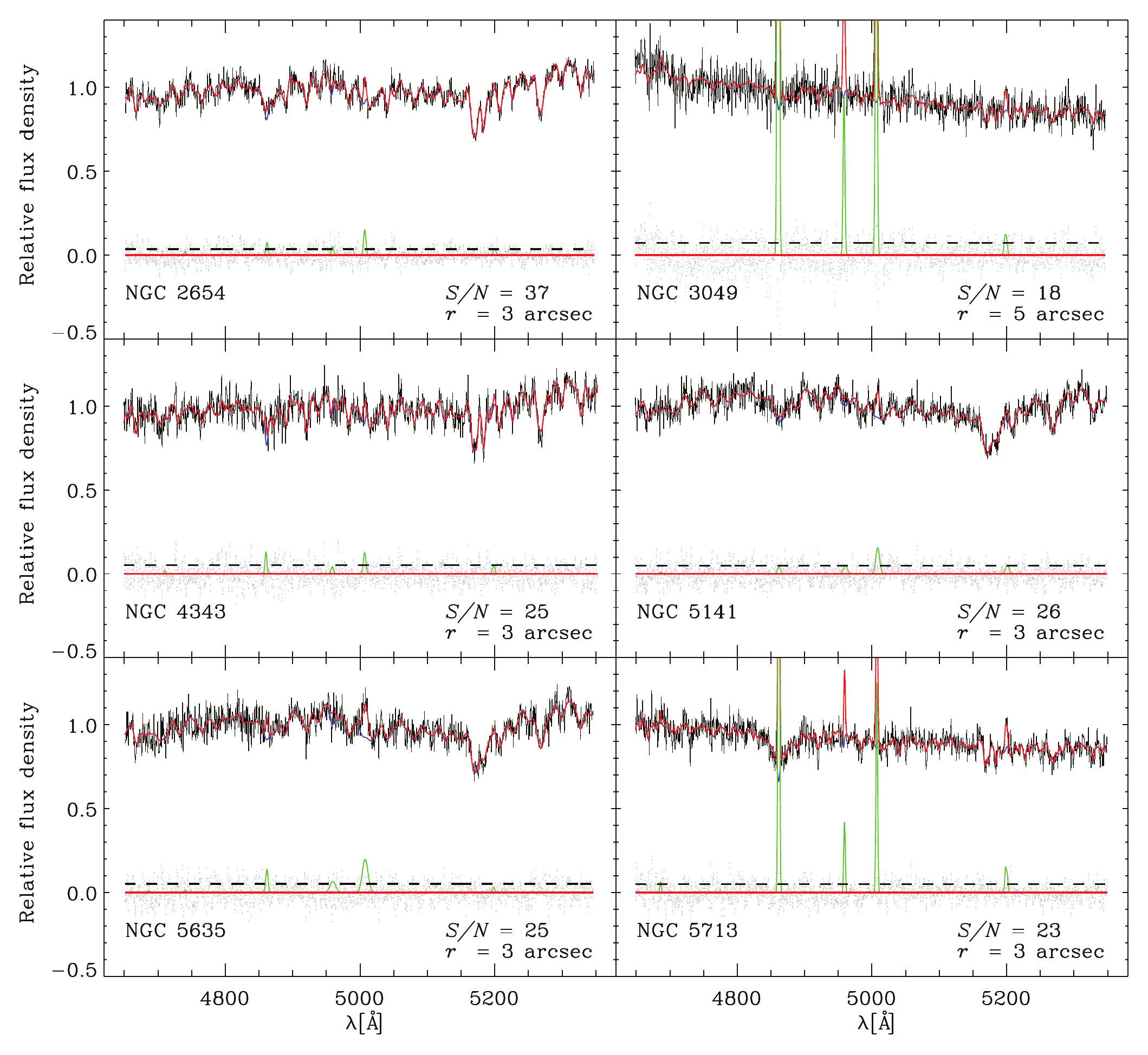}
\caption{As in Fig.~\ref{fig:bc}, but for the remaining sample galaxies.}
\label{fig:appT122}
\end{center}
\end{small}
\end{figure*}

\begin{figure*}[h]
\begin{small}
\begin{center}
\includegraphics[scale= 0.55,angle=0]{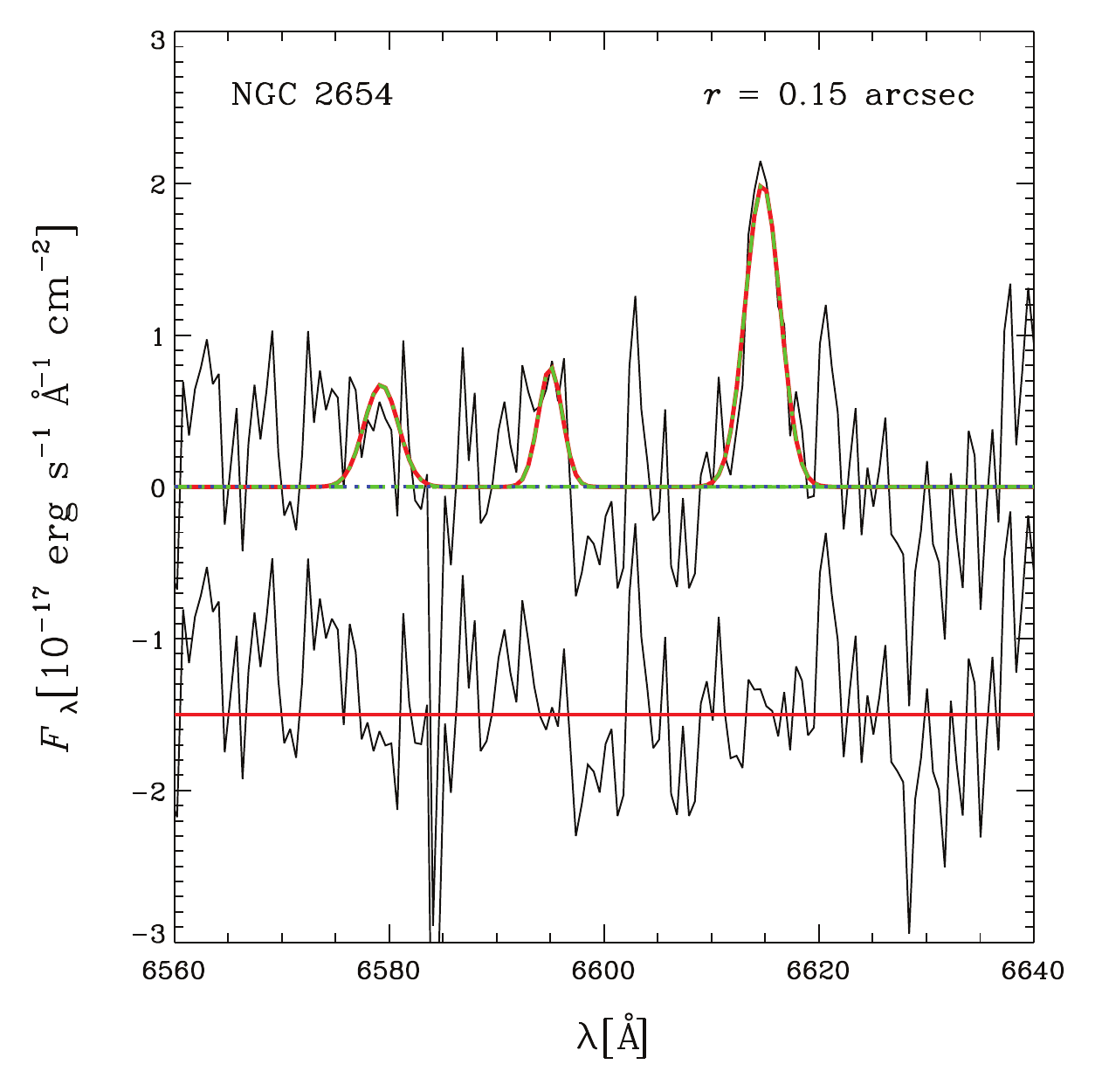}
\includegraphics[scale= 0.55,angle=0]{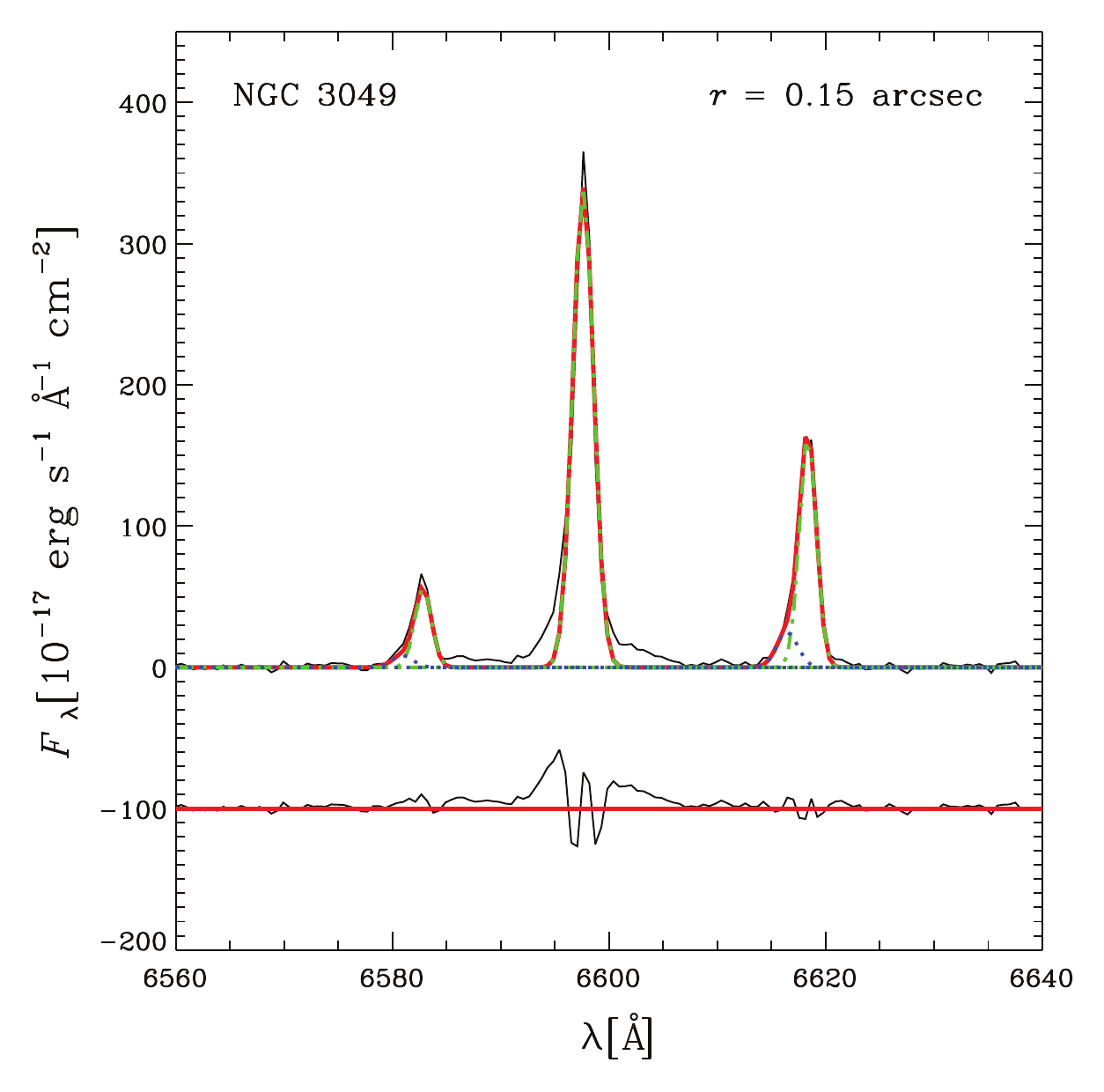}
\includegraphics[scale= 0.55,angle=0]{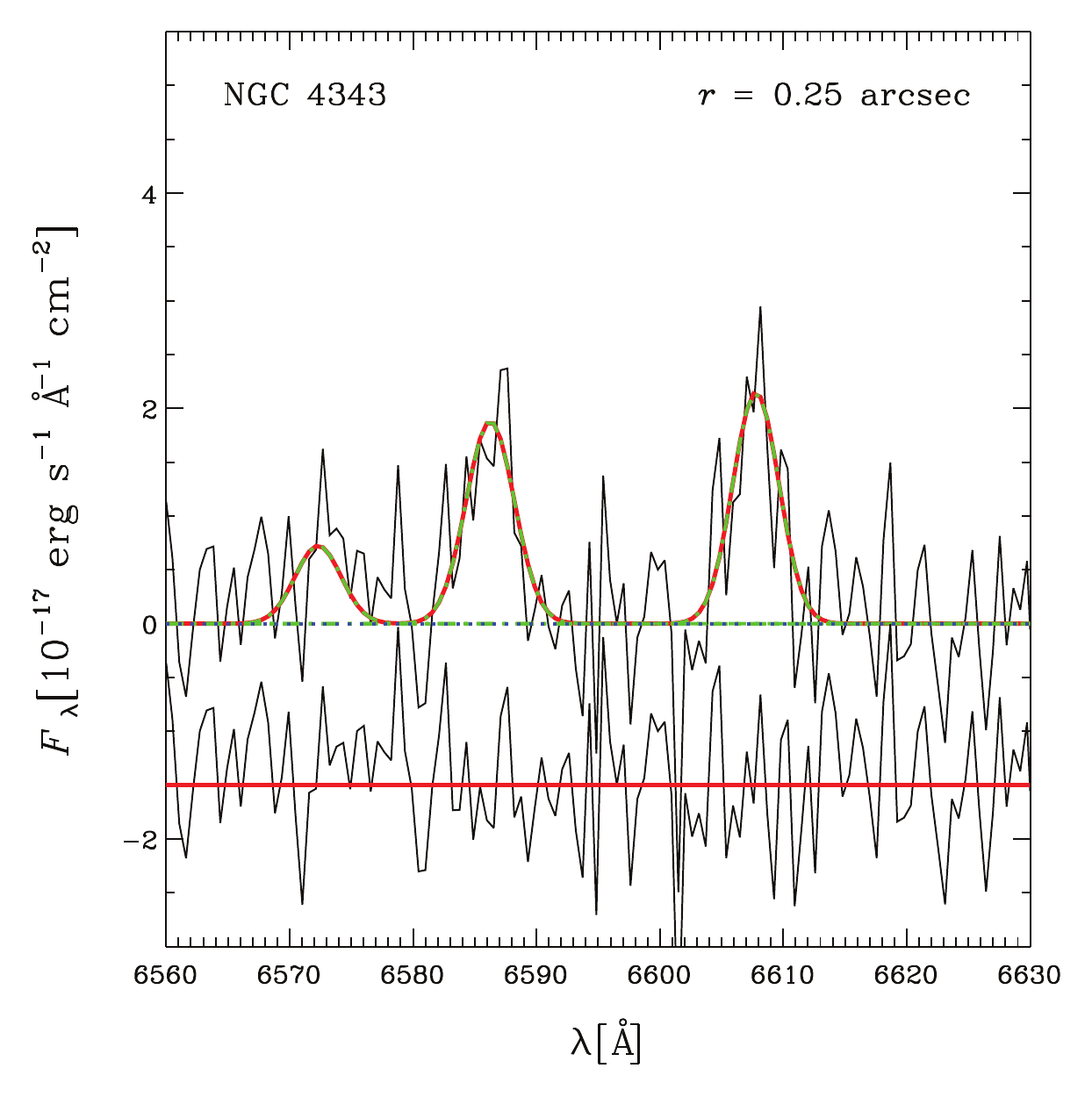}
\includegraphics[scale= 0.55,angle=0]{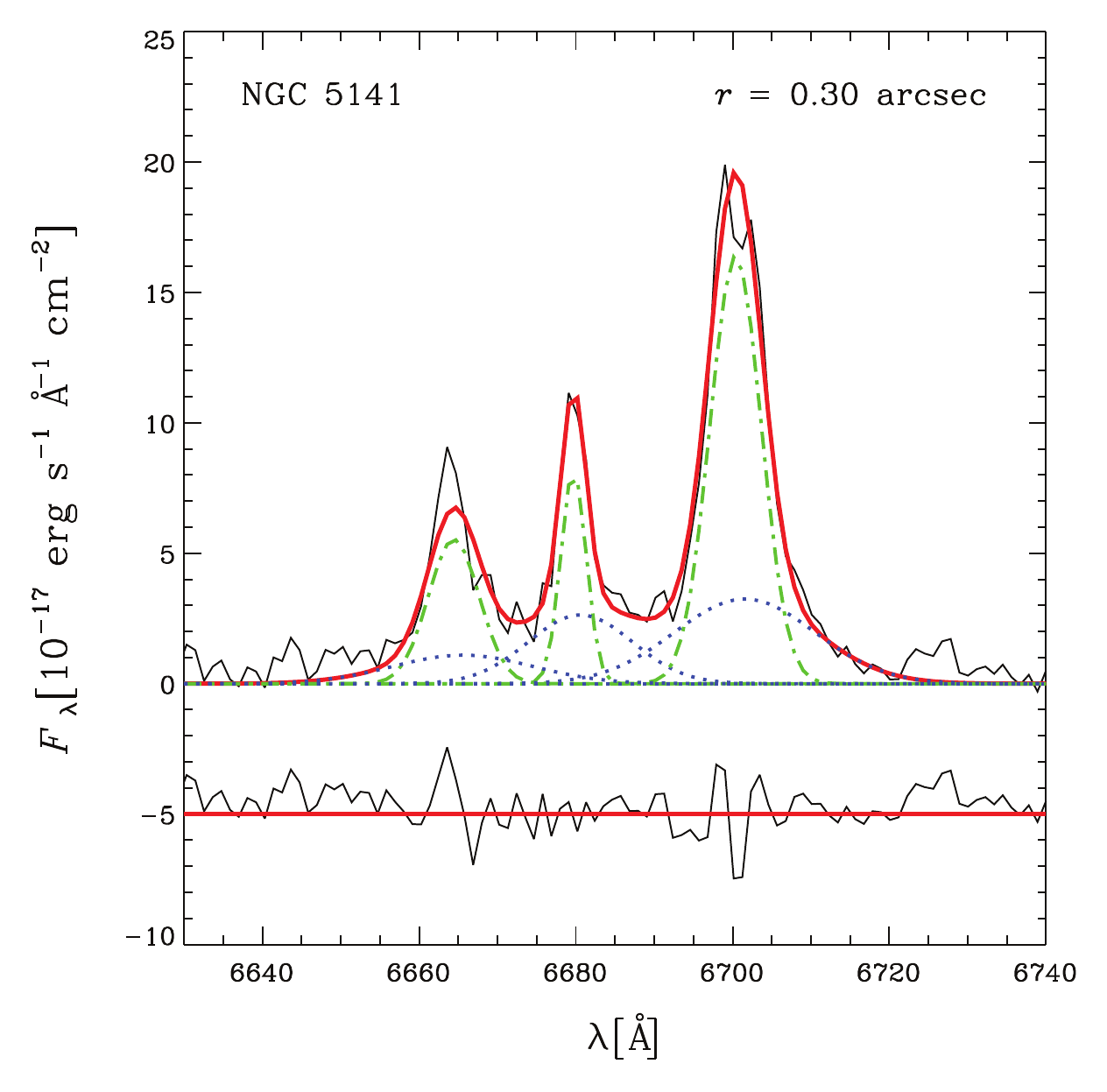}
\includegraphics[scale= 0.55,angle=0]{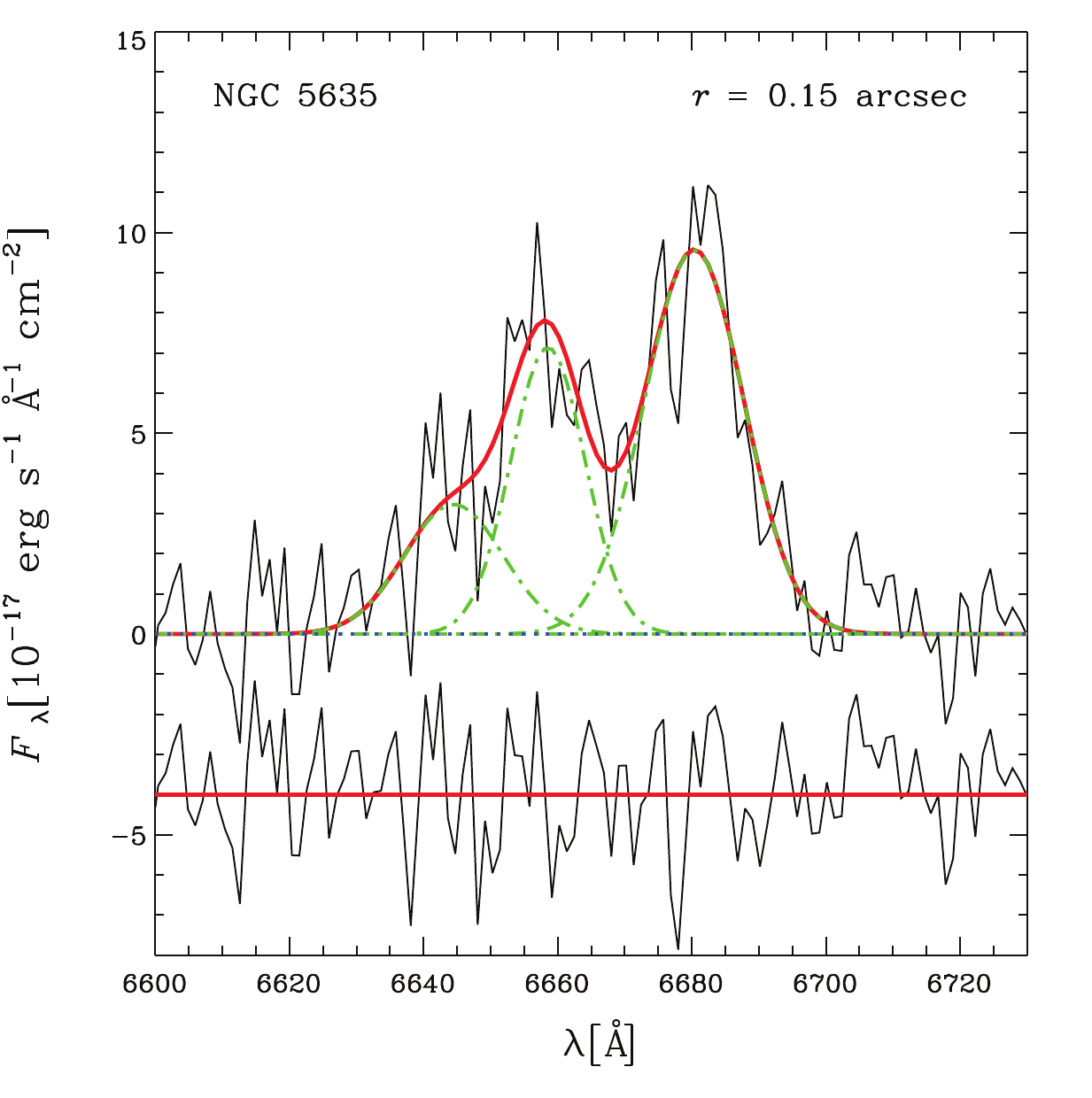}
\includegraphics[scale= 0.55,angle=0]{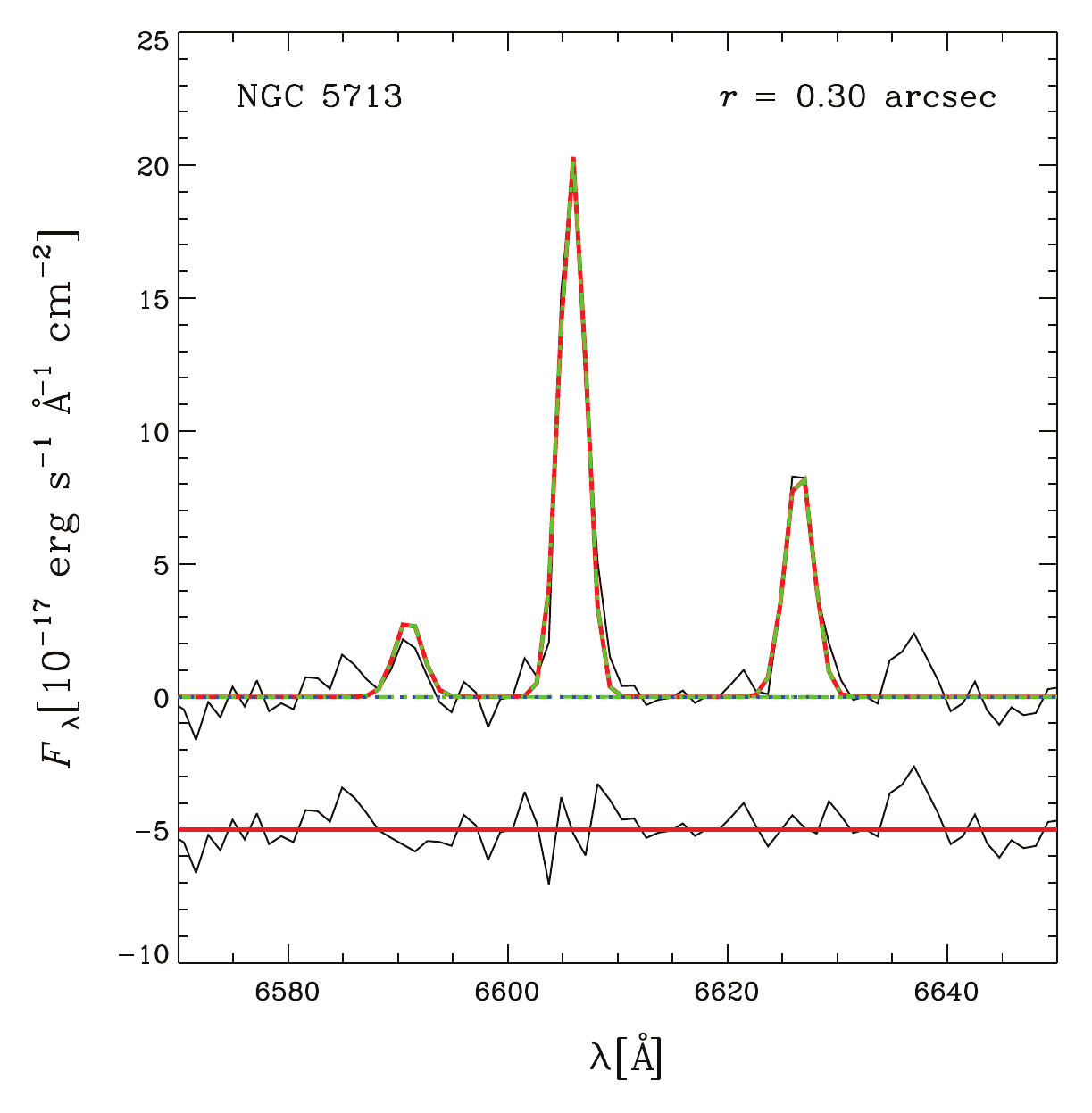}
\caption{As in Fig.~\ref{fig:stis}, but for the remaining sample galaxies.}
\label{fig:appstis}
\end{center}
\end{small}
\end{figure*}

\end{document}